\documentclass{ccjnl}

\usepackage{color}

\newcommand{\hem}{\mathrm{H}}
\newcommand{\tr}{\mathrm{Tr}}

\title{Achievable Rate Analysis and Optimization of Double-RIS Assisted Spatially Correlated MIMO with Statistical CSI}
\author{Kaizhe Xu\inst{1}, Jiajia Guo\inst{2}, Jun Zhang\inst{3}, Shi Jin\inst{4}, Shaodan Ma\inst{5,*}\corinfo{shaodanma@um.edu.mo}}
\receiveddate{}
\reviseddate{}
\Editor{}

\address[1]{Department of Communications and Networking, Xi'an Jiaotong-Liverpool University, Suzhou, 215123, China}
\address[2]{State Key Laboratory of Internet of Things for Smart City, University of Macau, Macao SAR, China}
\address[3]{Jiangsu Key Laboratory of Wireless Communications, Nanjing University of Posts and Telecommunications, Nanjing 210003, China}
\address[4]{National Mobile Communications Research Laboratory, Southeast University, Nanjing, 210096, China}
\address[5]{State Key Laboratory of Internet of Things for Smart City and the Department of Electrical and Computer Engineering, University of Macau, Macao SAR, China}

\begin{document}
\maketitle

\begin{abstract}
    Reconfigurable intelligent surface (RIS) is a novel meta-material
    which can form a smart radio environment by dynamically altering reflection directions of the impinging electromagnetic waves.
    In the prior literature, the inter-RIS links which also contribute to the performance of the whole system are usually neglected when multiple RISs are deployed.
    In this paper we investigate a general double-RIS assisted multiple-input multiple-output (MIMO) wireless communication system under spatially correlated non line-of-sight propagation channels, where the cooperation of the double RISs is also considered.
    The design objective is to maximize the achievable ergodic rate based on full statistical channel state information (CSI).
    Specifically, we firstly present a closed-form asymptotic expression for the achievable ergodic rate by utilizing replica method from statistical physics.
    Then a full statistical CSI-enabled optimal design is proposed which avoids high pilot training overhead compared to instantaneous CSI-enabled design. To further reduce the signal processing overhead and lower the complexity for practical realization, a common-phase scheme is proposed to design the double RISs.
    Simulation results show that the derived asymptotic ergodic rate is quite accurate even for small-sized antenna arrays. And the proposed optimization algorithm can achieve substantial gain at the expense of a low overhead and complexity. Furthermore, the cooperative double-RIS assisted MIMO framework is proven to achieve superior ergodic rate performance and high communication reliability under harsh propagation environment.
\end{abstract}

\section{introduction}
\label{s1}

The fifth generation (5G) wireless communication networks have been standardized and deployed worldwide \cite{6736746, 6798744}. However, the requirements for wireless communications keep ever-increasing, which has driven the research and exploration for next generation communication networks, i.e., 6G. In order to achieve extreme user experience, 6G must realize significant improvements in terms of key capabilities \cite{dang2020should, 9145564, 9925080}.
Based on the upgrade trends of earlier mobile generations, 6G is expected to provide universal high-performance wireless connections with tera-bits per second peak data rate, ten-fold increase in reliability, and at least 100 times energy efficiency, etc \cite{HuaweiSixG}.
Meanwhile, one of the key drivers and central topics for 6G is sustainability, particularly in terms of energy consumption of devices in the entire network. In this regard, the design of 6G infrastructure must lower the total power consumption and prioritize energy efficiency to make our planet a better place to live.
As a promising technology, reconfigurable intelligent surface (RIS) stood out as an energy-efficient solution for the next-generation wireless network \cite{8910627, tang20JSAC, 9969163}.
Generally speaking, RIS is a novel artificial electromagnetic meta-surface, which can manipulate the electromagnetic characteristics of each reflecting unit accurately and dynamically.
RIS is able to change the phase shifts of the impending electromagnetic wave thus forming a favourable radio propagation environment, leading the future wireless communication to a whole new spatial transmission paradigm \cite{9140329, 9136592}.

There have been numerous studies dedicated to exploiting the advantages of RIS to enhance transmission rate \cite{9110912, 9530675, 9343768, 9043523, 9279253, 9039554}.
Besides single-RIS assisted systems in the above works, multiple RISs have also been implemented to uplift the system performance to further increase the spatial degree of freedom (DoF) \cite{9497709, 9410435, 9448236, 9772371, 9729228}.
Under a spatially distributed multiple-RIS network, multiple RISs are organized to collaboratively achieve an energy efficient system in \cite{9497709}.
Since much heavier training overhead is required in multi-RIS deployment, \cite{9410435} utilized random beamforming and maximum likelihood estimation to obtain the optimal beam.
Beyond the link-level performance optimization, \cite{9448236} studied the RIS-user association design on network-level, where the average received signal-to-interference-plus-noise ratio (SINR) of each user is balanced to an optimal tradeoff. Futhermore, multiple RISs were also applied in the wideband mmWave MISO-OFDM systems \cite{9497353}.

However, most of the literature assumed that the transmitted signal was reflected by RIS only once, thus ignoring the inter-RIS cooperation \cite{mei2021multi, huang2021multi}. In practice, the inter-RIS link indeed exists as a significant link thus neglecting the second-hop reflection will lead to inaccurate modeling of the practical system.
In addition, the cooperation between RISs through second reflection provides another DoF to enhance the performance of RIS-assisted systems. It is proven that multiple RISs deployment with multi-hop reflection is able to realize a wilder range of smart radio environment coverage, more pronounced cooperative passive beamforming (CPB) gains, and higher path diversity \cite{dong2021double, han2020cooperative, 9247315, 9635819, 9852985, 9629293}. Therefore, the double-RIS assisted communications with secondary reflection have drawn certain attention recently.
The work \cite{9776512} investigated the impact of spatial correlation on the rate performance and a closed-form achievable rate expression was derived for the downlink double-RIS assisted multi-user MISO network. The influences of the double-RIS phase shifts on the average channel power was investigated in \cite{ding2022analysis}, where the performance of MISO system under several different Rician factors were discussed.
Besides the characterization of communication rate, transmission reliability was also evaluated.
The work \cite{9580460} presented the derivation of the coverage probability by applying a deterministic equivalent method for a double-RIS cooperatively assisted single-input-single-output (SISO) system. Nonetheless, these performance characterization of the SISO or MISO systems through scalar or vector operation can not be directly generalized to MIMO system. Due to the chain product of random matrices, a challenging random matrix theory (RMT) problem to characterize the successively cascaded channel of spatially correlated multi-hop RISs-assisted MIMO system has not been solved yet, which hinders the analysis of system performance.
With respect to the optimal design for double-RIS MIMO system, an efficient alternating optimization (AO) algorithm was proposed for the double-RIS system \cite{zheng2021double} to maximize the SINR,
where the perfect cascaded CSI was assumed available to design the transmission. Besides, the capacity maximization problem was also solved under line-of-sight (LoS) channel model in \cite{9714463}. By exploiting the characteristic of LoS channel structure, an algorithm with low complexity was proposed in order to optimize the two passive beamforming matrices.
It turns out that the double-RIS system enhances the system capacity by increasing the channel rank through adding the secondary reflection link and further increasing the spatial multiplexing gain.
However, more general non-LoS channel is not considered for double-RIS system in \cite{9714463}.

Nearly all the design schemes mentioned above assume perfect instantaneous CSI is known.
However, the acquisition of CSI is hard to realize due to the lack of radio frequency (RF) chains on the RISs, not to mention the surface generally contains a mass of channel coefficients to be estimated.
Moreover, the estimation of the
CSI between the RISs in a double-RIS system seems even more difficult \cite{9241706, zheng2021efficient}. Therefore, instantaneous CSI-enabled optimization is hard to realize in practice.
Without perfect instantaneous CSI, the robust design was studied based on the imperfect cascaded channel for BS-RIS-user link in \cite{9180053}.
Considering a more general system, \cite{9729228} studied the robust design of a general multi-RIS multi-user MIMO system. To enhance the system's global stability, statistical CSI errors were assumed as the uncertainty factor in \cite{9729228}.
Meanwhile, as another way to get rid of the dependence of perfect instantaneous CSI when conducting system optimization, statistical CSI-enabled design is a more feasible and efficient scheme \cite{9899454, 9690059, 9743440, 9992249}.
To reduce the substantial overhead caused by extremely high dimensions of estimated channels, \cite{9899454} derived a closed-form achievable weighted sum-rate expression under a cell-free system, then passive beamformers at RIS were optimized based on the statistical CSI.
Under a hybrid precoder framework, \cite{9690059} also conducted a statistical CSI-enabled RIS optimization through RIS reflection coefficient matrix relaxation.
However, the BSs' transmit beamforming still rely on the instantaneous CSI in \cite{9899454} and \cite{9690059} thus only partial statistical CSI are exploited.
Considering a more general multi-user MIMO system, \cite{9743440} and \cite{9992249} respectively investigated the uplink and downlink RIS-assisted system. Nevertheless, all these works deployed single RIS and did not consider the inter-RIS cooperation.
To enable the statistical CSI-enabled design for a double-RIS assisted MIMO system, a natural thinking is to find an accurate rate approximation dependent on slowly varying statistical parameters and take it as the objective for system design. However, no analytical rate expression dependent only on long-term statistics is available for a double-RIS assisted spatially correlated MIMO system currently.
The multi-hop relay assisted MIMO system has been studied previously \cite{fawaz2011asymptotic, hoydis2011iterative}, however, the conclusions cannot be applied to double-RIS assisted MIMO system directly due to the unique channel structure of double RIS.
In specific, signals impinging on the successive RISs will bounce off the surface immediately without generating any noise since no ADCs/DACs or amplifier is applied. Mathematically, double and single reflections will lead to a successive fading channel which is presented as a sum product of multiple matrices and is challenging to deal with.
\par
To take the advantages of distributed multiple-RIS framework and cooperative beamforming gain, in this work, we aim to analyze and maximize the achievable rate of the double-RIS assisted MIMO under Rayleigh fading, where full statistical CSI are exploited and the spatial correlations at the BS, the user, and the double RISs are also taken into consideration.
To the best of the authors' knowledge, there is no exiting work which has investigated such a practical system.

\subsection{Contributions}
In this paper, we analyze the ergodic rate and conduct statistical CSI-enabled optimal design, which unveils the potential benefit from the implementation of double RISs assisting the communication between a BS and a user. More specifically, the main contributions of this paper are highlighted below.
First, a double-RIS assisted MIMO system under spatially correlated Rayleigh fading is investigated. To conduct the ergodic rate analysis where a sophisticated matrices sum and product structure is involved, replica method in large-dimensional RMT is adopted to derive the achievable ergodic rate. Although the closed-form expression is nominally valid in the large system regime where the number of antennas tends to infinity, it is validated that the analytical results are also applicable in the case of a small number of antennas.

Second, given the derived closed-form asymptotic ergodic rate formula, an AO algorithm is proposed to conduct BS's active transmit beamforming and double RISs' passive beamforming in order to maximize the ergodic rate of the system.
By exploiting statistical CSI only, the proposed algorithm alternatively optimizes the source covariance matrix and the reflective phase-shifting matrices of the double RISs.
Specifically, we employ the water-filling algorithm for the source covariance matrix design. For the phase-shifting matrices design of the double RISs, the particle swarm optimization (PSO) algorithm is utilized to obtain a local optima. To further reduce complexity, a common-phase scheme is also proposed, where the double RISs utilize equal reflecting matrices.

Third, we present the numerical simulations result to validate the accuracy of the derived closed-form asymptotic ergodic rate, the efficiency of the proposed AO algorithm, and the advantages of the common-phase scheme.
Also the benefit from the double-RIS MIMO framework is investigated.
The joint optimization problem is solved by exploiting full statistical CSI based on the asymptotic analysis of the double-RIS assisted MIMO under common reflection pattern.
This indicates that our design alleviates high computational complexity and avoids the tedious CSI acquisition process in the double-RIS assisted MIMO communication system.  
Furthermore, with the inter-RIS transmission link the double-RIS assisted MIMO system can achieve a better ergodic rate performance and higher system reliability than that without considering the double reflection between the RISs.

\subsection{Outline}
In Section II, the considered double-RIS assisted MIMO system, channel models, and problem formulation are described. In Section III, the asymptotic ergodic rate expression is derived under the large-system regime. The optimization problem is solved in Section IV by proposing our AO algorithm and common-phase scheme. In Section V, numerical results are presented. Section VI concludes the paper while the Appendix sketches the proof of the asymptotic results by RMT method.

\subsection{Notations}
In this paper, scalars, vectors, and matrices are respectively denoted by non-boldface, lowercase boldface, and uppercase boldface letters. $ {\bf 1} $, $ {\bf 0} $, and $ {\bf I}_N $ respectively denote all-one matrix with defined dimension, vector with all-zero elements, and $ N \times N $ identity matrix.
Notations $ \det(\cdot) $, $ (\cdot)^{-1} $, $ (\cdot)^\hem $, $ \text{vec}(\cdot) $, $ \tr (\cdot) $, and $ (\cdot)^{-\frac{1}{2}} $ denote respectively determinant, inverse, Hermitian, vectorization, trace, and matrix principal square root.
$ \mathbb{E} \lbrace \cdot \rbrace $ denotes expectation operator. And $ \mathbb{C} $ denotes the complex number set.
The natural logarithm and the Kronecker product are respectively denoted by $ \log $ and $ \otimes $.
Tabel \ref{tab1} summarizes the main symbol notations.

\begin{table}[!ht]
    \centering
    \caption{Main Symbol Notations.}
    \label{tab1}
    \begin{tabular}{cc}\toprule
    Symbol & Description \\\midrule
    $ M $, $ N $, $ L_i $ & Numbers of antennas at corresponding nodes \\
    $ {\bf H}_j $ & Channel matrices \\
    $ {\bf Q} $, $ {\bf \Theta}_i $ & Transmit covariance matrix and phase-shifting matrices \\
    $ {\bf T}_j $, $ {\bf R}_j $ & Transmit and receive correlation matrices \\
    $ {\bf W}_j $ & Rayleigh fading matrices \\
    $ {\bf \Gamma}_j $ & Large-scale fading coefficients \\
    $ R $, $ {\bar R} $ & Ergodic rate and asymptotic ergodic rate \\
    $ P $ & Maximum transmit power \\
    $ \{ e_j, {\tilde e}_j \} $ & Auxiliary variables in asymptotic result \\
    $ \iota $, $ {\bf \Upsilon} $ & Lagrange multipliers \\
    $ {\bf u}_m $ & 3-D Cartesian coordinations of RISs' elements \\
    $ F $ & Free energy \\
    $ {\bf V}_j $ & Replica sets \\
    $ {\bf C}_j $ & Covariance matrices of random channel matrices \\
    $ {\mathcal G}_j^{(r)} $, $ {\mu}_j^{(r)} $ & Measures of ${\bf C}_j$\\
    $ {\mathcal R}_j^{(r)} $ & Rate measure of ${\bf C}_j$\\\bottomrule
    \end{tabular}
    \end{table}

\section{System Model And Problem Formulation}
\begin{figure}[!ht]
    \centering
    \includegraphics[width=0.52\textwidth]{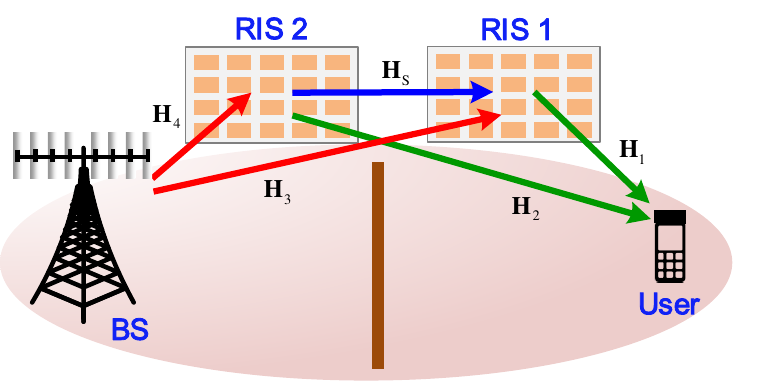}
    \caption{A cooperatively double-RIS assisted MIMO system.}
    \label{systemmodel_double_RIS}
    \end{figure}

In this paper, a double-RIS assisted MIMO wireless communication system is considered, where two RISs are deployed to assist one base station (BS) to serve one user, as shown in Fig. \ref{systemmodel_double_RIS}. Due to the unfavorable propagation environment, the direct link from the BS and the user is blocked by substantial obstructions \cite{huang2019reconfigurable, 9133142}, thus the data transmission from the BS to the user can only be completed by the double-RIS reflection links established by the $ 2 $ RISs thanks to the coverage extension capability of RISs. It is worth noting that we can also extend the result in this work to the scenario with a direct link, only with one more random channel matrix to be considered and higher computational complexity. In order to minimize the effective path loss, one RIS is deployed closed to the BS and the other RIS is deployed in the vicinity of the user \cite{zhang2021large}. The BS, the user, and RIS $ i $ are implemented with $ M $ antennas, $ N $ antennas, and $ L_i, i \in \{ 1, 2 \} $ reflecting elements respectively.
The signals transmitted from the BS impinge on the two RISs respectively through the channels between the BS and RIS $ i $, then the signals are reflected to the user from the two RISs. Especially, there exists a secondary reflection link between the two RISs, which forms a three-hop end-to-end channel. $ {\bf H}_1 \in \mathbb{C}^{N \times L_1}, {\bf H}_2 \in \mathbb{C}^{N \times L_2}, {\bf H}_3 \in \mathbb{C}^{L_1 \times M}, {\bf H}_4 \in \mathbb{C}^{L_2 \times M} $ denote respectively the channel matrices of RIS1-user, RIS2-user, BS-RIS1, BS-RIS2 channels. It is worth noting that the indexes of the RISs and the channels are assigned from near to far from the user, which is convenient for the following derivation. To highlight the particularity of the inter-RIS channel, we denote RIS2-RIS1 channel as $ {\bf H}_{\rm s} \in \mathbb{C}^{L_1 \times L_2} $.
The square reflective phase-shifting matrix for RIS $ i $ is denoted by $ {\bf \Theta}_i = {\rm diag} \{ e^{j \theta_{i, 1}}, e^{j \theta_{i, 2}}, \cdots, e^{j \theta_{i, {L_i}}} \} \in \mathbb{C}^{L_i \times L_i} $.
Each reflecting element $ {\vartheta}_{i,l} = e^{j \theta_{i,l}}, ~\forall i,l $ obeys unit-modular constraints $ \vert {\vartheta}_{i,l} \vert = 1 $ whose phases are $ \theta_{i,l} \in \lbrack 0, 2\pi) $.
Therefore, the effective channel from the BS to the user can be expressed as
    \begin{align}
    \label{eff_Ch}
    {\bf H}_{\rm {eff}} = {\bf H}_1 {\bf \Theta}_1 {\bf H}_3 + {\bf H}_2 {\bf \Theta}_2 {\bf H}_4 + {\bf H}_1 {\bf \Theta}_1 {\bf H}_{\rm s} {\bf \Theta}_2 {\bf H}_4,
	\end{align}
where the first two terms each including a two cascaded channels structure are corresponding to the two single reflection links, and the third term including a three cascaded channels structure is corresponding to the double reflection link. Denote $ {\bf x} \in \mathbb{C}^{M} $ a zero mean Gaussian transmitting signal vector with covariance matrix $ {\bf Q} \in \mathbb{C}^{M \times M} $, $ {\bf n} \in \mathbb{C}^{N} $ a zero mean Gaussian noise vector whose covariance matrix is $ \sigma^2 {\bf I}_{N} $. Then, the signal received at the user
can be expressed as
    \begin{align}
    \label{received_signal}
	{\bf y} = {\bf H}_{\rm {eff}} {\bf x} + {\bf n}.
	\end{align}
To characterize the MIMO channel with spatial correlations for each link, we adopt Kronecker model which can be written as ${\bf H}_j = {\bf R}_j^{\frac {1}{2}} {\bf W}_j {\bf T}_j^{\frac {1}{2}}, j \in \{ 1, 2, 3, 4, \rm{s} \}$,  where
$ {\bf R}_1 \in {\mathbb C}^{ N \times N }, {\bf R}_2 \in {\mathbb C}^{ N \times N }, {\bf R}_3 \in {\mathbb C}^{ L_1 \times L_1 }, {\bf R}_4 \in {\mathbb C}^{ L_2 \times L_2 }, {\bf R}_{\rm s} \in {\mathbb C}^{ L_1 \times L_1 } $, and $ {\bf T}_1 \in {\mathbb C}^{ L_1 \times L_1 }, {\bf T}_2 \in {\mathbb C}^{ L_2 \times L_2 }, {\bf T}_3 \in {\mathbb C}^{ M \times M }, {\bf T}_4 \in {\mathbb C}^{ M \times M }, {\bf T}_{\rm s} \in {\mathbb C}^{ L_2 \times L_2 } $ are deterministic nonnegative semi-definite receive and transmit correlation matrices at the corresponding nodes. $ {\bf W}_1 \equiv [ \frac{1}{\sqrt{L_1}} W_{1, ab} ] \in {\mathbb C}^{ N \times L_1 }, {\bf W}_2 \equiv [ \frac{1}{\sqrt{L_2}} W_{2, ab} ] \in {\mathbb C}^{ N \times L_2 }, {\bf W}_3 \equiv [ \frac{1}{\sqrt{M}} W_{3, ab} ] \in {\mathbb C}^{ L_1 \times M }, {\bf W}_4 \equiv [ \frac{1}{\sqrt{M}} W_{4, ab} ] \in {\mathbb C}^{ L_2 \times M }, {\bf W}_{\rm s} \equiv [ \frac{1}{\sqrt{L_2}} W_{{\rm s}, ab} ] \in {\mathbb C}^{ L_1 \times L_2 } $ consist of random components where the elements $ W_{i, ab} $'s are independent and identically distributed (i.i.d.) complex zero-mean unit-variance random variables$^1$.
{\footnotetext[1]{Results under Rician fading can also be extended from our results in this paper, where a line-of-sight component is involved and higher computational complexity is introduced.}}
Furthermore, the large-scale fading coefficients of all the links are denoted by $ \Gamma_j, ~\forall j $. For the deterministic terms, $ {\bf R}_j, {\bf T}_j $ as normalized below,
    \begin{subequations}
    \begin{align}
    & \tr({\bf R}_1) = N, \ \tr({\bf T}_1) = L_1^2 \Gamma_1, \\
    & \tr({\bf R}_2) = N, \ \tr({\bf T}_2) = L_2^2 \Gamma_2, \\
    & \tr({\bf R}_{\rm s}) = L_1, \ \tr({\bf T}_{\rm s}) = L_2^2 \Gamma_{\rm s}, \\
    & \tr({\bf R}_3) = L_1, \ \tr({\bf T}_3) = M \Gamma_3, \\
    & \tr({\bf R}_4) = L_2, \ \tr({\bf T}_4) = M \Gamma_4.
    \end{align}
    \end{subequations}
With the source covariance matrix $ {\bf Q} $ and the controllable phase-shifting matrices $ {\bf \Theta}_i, i \in \{ 1, 2 \} $, for the successively cascaded MIMO channel, the achievable ergodic rate is expressed as
    \begin{align}
    \label{R}
    R({\bf Q}, {\bf \Theta}_i) =  \mathbb{E}_{ \{ {\bf H}_j \} }
    \left\{ \log \det \left( {\bf I}_N + \frac{1}{\sigma^2} {{\bf H}_{ \rm {eff} }} {\bf Q} {{\bf H}_{ \rm {eff} }^\hem} \right) \right\}.
    \end{align}
Our objective is to maximize the achievable ergodic rate by a combined optimum design of the source covariance matrix $ {\bf Q} $ and the controllable phase-shifting matrices $ {\bf \Theta}_i, i \in \{ 1, 2 \} $. With the power constraints on the BS and the unit-modulus constraints on the passive beamforming of the RISs, the optimization problem is formulated as
    \begin{align}
    \label{P1}
    \text{(P1)} \quad & \max_{ {\bf Q}, \boldsymbol{\Theta}_1, \boldsymbol{\Theta}_2 } \ \ R\left( {{\bf{Q}}, {\bf{\Theta }}_1, {\bf{\Theta }}_2} \right) \\
    \text{s.t.} \ \ \ & {\bf \Theta}_i = {\rm diag} \{ \vartheta_{i, 1}, \vartheta_{i, 2}, \cdots, \vartheta_{i, L_i} \}, i \in \{1, 2\}, \\
    & \lvert \vartheta_{i, l_i} \rvert = 1,  1 \le l_i \le L_i, \\
    & \tr\left( {\bf Q} \right) \le P, {\bf Q} \succeq 0.
    \end{align}
Obviously, the challenges of solving optimization problem (P1) are mainly three-fold. Firstly, both the unit-modulus constraint and the diagonal structure of phase-shifting matrices make problem (P1) non-convex. Secondly, the three optimization variables $ \{ {\bf{Q}}, {\bf{\Theta }}_1, {\bf{\Theta }}_2 \} $ are coupling with each others. Finally, the calculation of achievable ergodic rate involves high-dimensional integrals, which requires Monte-Carlo simulation with prohibitively computational complexity. In the following, we propose an analysis method in large system regime to find the optimal solutions for the double-RIS assisted MIMO transceiver design.

\section{Large System Analysis}
In this section, we propose a large system analysis to obtain the asymptotic achievable rate, where it is assumed that the antennas number tends to infinity while keeping the ratios constant, i.e., $ L_1 / M = \eta_1 $, $ L_2 / M = \eta_2 $, $ N / L_1 = \eta_3 $, $ N / L_2 = \eta_4 $. The effects of of the two optimization variables, i.e., $ {\bf Q} $ and $ {\bf \Theta}_i $ are absorbed into $ \{ {\bf T}_j, ~\forall j \} $ to simplify the derivation by the following replacements
    \begin{subequations}
    \label{replacement}
    \begin{align}
    & {\bf T}_1^{{\bf \Theta}_1} := {\bf \Theta}_1^\hem {\bf T}_1 {\bf \Theta}_1, {\bf T}_2^{{\bf \Theta}_2} := {\bf \Theta}_2^\hem {\bf T}_2 {\bf \Theta}_2, \\
    & {\bf T}_{\rm s}^{{\bf \Theta}_2} := {\bf \Theta}_2^\hem {\bf T}_{\rm s} {\bf \Theta}_2, \\
    & {\bf T}_3^{{\bf Q}} := {\bf Q}^\frac{1}{2} {\bf T}_3 {\bf Q}^\frac{1}{2}, {\bf T}_4^{{\bf Q}} := {\bf Q}^\frac{1}{2} {\bf T}_4 {\bf Q}^\frac{1}{2}.
    \end{align}
    \end{subequations} 
    For ease of notation, the superscript $ {\bf Q} $ and $ {\bf \Theta}_i $ in $ \{ {\bf T}_j^{{\bf \Theta}_i}, {\bf T}_j^{{\bf Q}}, ~\forall j \} $ are omitted during the analysis derivation, and simply expressed as $ \{ {\bf T}_j, ~\forall j \} $. Then the effective channel can be rewritten as
    \begin{align}
    {\bf H}_{\rm {eff}} & := {\bf H}_1 {\bf H}_3 + {\bf H}_2 {\bf H}_4 + {\bf H}_1 {\bf H}_{\rm s} {\bf H}_4 \\ \notag
    & = ({{\bf{H}}_1},\;{{\bf{H}}_2})\left( {\begin{array}{*{20}{c}}
    {{{\bf{I}}_{{L_1}}}}&{{{\bf{H}}_{\rm{s}}}}\\
    {{{\bf{0}}_{{L_2} \times {L_1}}}}&{{{\bf{I}}_{{L_2}}}}
    \end{array}} \right) \left({\bf H}_3^\hem, {\bf H}_4^\hem \right)^\hem \\ \notag
    & = {{\bf{H}}_{\rm{r}}}{{{\bf{\mathord{\buildrel{\lower3pt\hbox{$\scriptscriptstyle\frown$}}
\over H} }}}_{\rm{s}}}{{\bf{H}}_{\rm{t}}},
    \end{align}
where the channel matrices are compacted as effective receiving channel matrix $ {\bf H}_{\rm r} = ({\bf H}_1, {\bf H}_2) \in {\mathbb C}^{ N \times (L_1 + L_2) } $, effective inter-RIS reflecting channel matrix $ {{{\bf{\mathord{\buildrel{\lower3pt\hbox{$\scriptscriptstyle\frown$}}
\over H} }}}_{\rm{s}}} = \left( {\begin{array}{*{20}{c}}
{{{\bf{I}}_{{L_1}}}}&{{{\bf{H}}_{\rm{s}}}}\\
{{{\bf{0}}_{{L_2} \times {L_1}}}}&{{{\bf{I}}_{{L_2}}}}
\end{array}} \right) \in {\mathbb C}^{ (L_1 + L_2) \times (L_1 + L_2) } $, and effective transmitting channel matrix $ {\bf H}_{\rm t} = ({\bf H}_3^\hem, {\bf H}_4^\hem)^\hem \in {\mathbb C}^{ (L_1 + L_2) \times M } $. Therefore, the achievable ergodic rate $ R $ in (\ref{R}) can be re-expressed by a compacted formula
    \begin{align}
    \label{ergodic_rate_after_replacement}
    & I = \mathbb{E}_{\lbrace {\bf H}_j \rbrace}
    {\lbrace \log \det ( {\bf I}_N + \frac{1}{\sigma^2} ( {{\bf{H}}_{\rm{r}}}{{{\bf{\mathord{\buildrel{\lower3pt\hbox{$\scriptscriptstyle\frown$}}
\over H} }}}_{\rm{s}}}{{\bf{H}}_{\rm{t}}} ) ( {{\bf{H}}_{\rm{r}}}{{{\bf{\mathord{\buildrel{\lower3pt\hbox{$\scriptscriptstyle\frown$}}
\over H} }}}_{\rm{s}}}{{\bf{H}}_{\rm{t}}} )^\hem ) \rbrace}.
    \end{align}
The sophisticated matrices sum and product structure in the effective channel throws many challenges in deriving the asymptotic achievable ergodic rate. First, different from the SISO and MISO systems, characterizing the double-RIS assisted MIMO system involves matrices operations which are far more complicated than those of scalar and vector. The analysis of the achievable ergodic rate in (\ref{ergodic_rate_after_replacement}) is essentially a RMT problem. Second, the inter-RIS reflection link $ {\bf H}_{\rm s} $ is considered in our model, previous results of RIS-assisted MIMO systems considering only single reflections are not applicable \cite{zhang2021large}. This is because the effective channel (\ref{eff_Ch}) not only involves two matrices multiplications but also a three matrices multiplication. No analytical result, particularly for the product of three random matrices, is currently available for our system.
To address these issues, we resort to replica method to calculate the high-dimensional integral to obtain the asymptotic achievable ergodic rate for double-RIS assisted MIMO system. Different from the previous work \cite{zhang2021large, 9530675}, we creatively form replica set with novel form according to the novel matrices sum and product structure. We obtain {\bf {Proposition 1}} as follows under the large system assumption: \\
    \begin{proposition}
    \label{proposition1}
    The asymptotic achievable ergodic sum-rate $ I $ for double-RIS assisted MIMO in (\ref{ergodic_rate_after_replacement}) can be given by
    \begin{align}
    \label{asy_I}
    {\bar I} = & \log \det \left( {\frac{{{\sigma ^2}{\bf{I}} + {e_1}{{\bf{R}}_1} + {e_2}{{\bf{R}}_2}}}{{{\sigma ^2}}}} \right) \\ \notag
    & + \log \det \left( {\bf{I}} + {{\tilde e}_1}{{\bf{T}}_1} ({e_{\rm{s}}}{{\bf{R}}_{\rm{s}}} + e_3 {\bf R}_3) \right) \\ \notag
    & + \log \det \left( {{\bf{I}} + \left( {{{\tilde e}_{\rm{s}}}{{\bf{T}}_{\rm{s}}} + {{\tilde e}_2}{{\bf{T}}_2}} \right){e_4}{{\bf{R}}_4}} \right) \\ \notag
    & + \log \det \left( {{\bf{I}} + {{\tilde e}_{\rm{3}}}{{\bf{T}}_3} + {{\tilde e}_{\rm{4}}}{{\bf{T}}_4}} \right) \\ \notag
    & - {L_1}{e_1}{{\tilde e}_1} - {L_2}{e_2}{{\tilde e}_2} - {L_2}{e_{\rm{s}}}{{\tilde e}_{\rm{s}}} - M{e_3}{{\tilde e}_{\rm{3}}} - M{e_4}{{\tilde e}_{\rm{4}}}, \notag
    \end{align}
    where the auxiliary variables are the unique solutions of the following system of equations$^{1, 2}$
{\footnotetext[1]{Due to the tedious expressions of the equation system, the dimension indicators of identity matrices and the parantheses following Tr are omitted.}}
{\footnotetext[2]{By using an iterative algorithm, we can obtain the unique solutions of $ \{ e_j, {\tilde e}_j, ~\forall j \} $. Since the iteration process is discussed in \cite{zhang2021large}, we omit the proof of the existence and uniqueness of the solution.}}
\begin{subequations}
    \label{auxiliary_variable_e}
    \begin{align}
    & {e_1} = \frac{1}{{{L_1}}}{\rm{Tr}}\;{{\bf{\Psi }}^{ - 1}}{{\bf{T}}_1}{e_{\rm{s}}}{{\bf{R}}_{\rm{s}}} \\ \notag
    & + {\left( {{\bf{I}} + \left( {{\bf{I}} - {{\tilde e}_1}{{\bf{T}}_1}{e_{\rm{s}}}{{\bf{R}}_{\rm{s}}}{{\bf{\Psi }}^{ - 1}}} \right){{\tilde e}_1}{{\bf{T}}_1}{e_3}{{\bf{R}}_3}} \right)^{ - 1}} \\ \notag
    & \times ( ( {{\bf{I}} - {{\tilde e}_1}{{\bf{T}}_1}{e_{\rm{s}}}{{\bf{R}}_{\rm{s}}}{{\bf{\Psi }}^{ - 1}}} ){{\bf{T}}_1}{e_3}{{\bf{R}}_3} \\ \notag
    & - {{\bf{T}}_1}{e_{\rm{s}}}{{\bf{R}}_{\rm{s}}}{{\bf{\Psi }}^{ - 1}}{{\tilde e}_1}{{\bf{T}}_1}{e_3}{{\bf{R}}_3} ), \\
    & {{\tilde e}_1} = \frac{1}{{{L_1}}}{\rm{Tr}}\;{\left( {{\sigma ^2}{\bf{I}} + {e_1}{{\bf{R}}_1} + {e_2}{{\bf{R}}_2}} \right)^{ - 1}}{{\bf{R}}_1},\\
    & {e_2} = \frac{1}{{{L_2}}}{\rm{Tr}}\;{\left( {{\bf{I}} + \left( {{{\tilde e}_{\rm{s}}}{{\bf{T}}_{\rm{s}}} + {{\tilde e}_2}{{\bf{T}}_2}} \right){e_4}{{\bf{R}}_4}} \right)^{ - 1}}{{\bf{T}}_2}{e_4}{{\bf{R}}_4},\\
    & {{\tilde e}_2} = \frac{1}{{{L_2}}}{\rm{Tr}}\;{\left( {{\sigma ^2}{\bf{I}} + {e_1}{{\bf{R}}_1} + {e_2}{{\bf{R}}_2}} \right)^{ - 1}}{{\bf{R}}_2},\\
    & {e_{\rm{s}}} = \frac{1}{{{L_2}}}{\rm{Tr}}\;{\left( {{\bf{I}} + \left( {{{\tilde e}_{\rm{s}}}{{\bf{T}}_{\rm{s}}} + {{\tilde e}_2}{{\bf{T}}_2}} \right){e_4}{{\bf{R}}_4}} \right)^{ - 1}}{{\bf{T}}_{\rm{s}}}{e_4}{{\bf{R}}_4},\\
    & {{\tilde e}_{\rm{s}}} = \frac{1}{{{L_2}}}{\rm{Tr}}\;{{\bf{\Psi }}^{ - 1}}{{\tilde e}_1}{{\bf{T}}_1}{{\bf{R}}_{\rm{s}}} \\ \notag
    & + {\left( {{\bf{I}} + \left( {{\bf{I}} - {{\tilde e}_1}{{\bf{T}}_1}{e_{\rm{s}}}{{\bf{R}}_{\rm{s}}}{{\bf{\Psi }}^{ - 1}}} \right){{\tilde e}_1}{{\bf{T}}_1}{e_3}{{\bf{R}}_3}} \right)^{ - 1}} {\bf \Xi},\\
    & {e_{\rm{3}}}{\rm{ = }}\frac{1}{M}{\rm{Tr}}\;{\left( {{\bf{I}} + \left( {{{\tilde e}_{\rm{3}}}{{\bf{T}}_3} + {{\tilde e}_{\rm{4}}}{{\bf{T}}_4}} \right)} \right)^{ - 1}}{{\bf{T}}_3},\\
    & {{\tilde e}_3} = \frac{1}{M}{\rm{Tr}}\;{\left( {{\bf{I}} + \left( {{\bf{I}} - {{\tilde e}_1}{{\bf{T}}_1}{e_{\rm{s}}}{{\bf{R}}_{\rm{s}}}{{\bf{\Psi }}^{ - 1}}} \right){{\tilde e}_1}{{\bf{T}}_1}{e_3}{{\bf{R}}_3}} \right)^{ - 1}} \\ \notag
    & \times \left( {{\bf{I}} - {{\tilde e}_1}{{\bf{T}}_1}{e_{\rm{s}}}{{\bf{R}}_{\rm{s}}}{{\bf{\Psi }}^{ - 1}}} \right){{\tilde e}_1}{{\bf{T}}_1}{{\bf{R}}_3},\\
    & {e_{\rm{4}}}{\rm{ = }}\frac{1}{M}{\rm{Tr}}\;{\left( {{\bf{I}} + \left( {{{\tilde e}_{\rm{3}}}{{\bf{T}}_3} + {{\tilde e}_{\rm{4}}}{{\bf{T}}_4}} \right)} \right)^{ - 1}}{{\bf{T}}_4}, \\
    & {{\tilde e}_4}{\rm{ = }}\frac{1}{M}{\rm{Tr}}\;{\left( {{\bf{I}} + \left( {{{\tilde e}_{\rm{s}}}{{\bf{T}}_{\rm{s}}} + {{\tilde e}_2}{{\bf{T}}_2}} \right){e_4}{{\bf{R}}_4}} \right)^{ - 1}}{{\bf{\Phi }}_2}{{\bf{R}}_4},
    \end{align}
    \end{subequations}
    with $ {\bf{\Psi }} = {\bf{I}} + {\tilde e_1}{{\bf{T}}_1}{e_{\rm{s}}}{{\bf{R}}_{\rm{s}}} $ and $ {\bf \Xi} = { - {{\tilde e}_1}{{\bf{T}}_1}{{\bf{R}}_{\rm{s}}}{{\bf{\Psi }}^{ - 1}}{{\tilde e}_1}{{\bf{T}}_1}{e_3}{{\bf{R}}_3}} $.       
    \end{proposition}
{\emph {Proof:}} See Appendix.\\
Notice that the analysis of the product of three random matrices is rarely investigated in the literature, and our analytical result can also serve as a good reference for other analysis involving multiple random matrices multiplication, which may have a wide range of applications in multihop networks  and multi-RIS-aided systems. In addition, we have the following insights.
    \begin{remark}
    \label{remark1}
    The result in { \bf{Proposition 1} } corresponding to the general double-RIS system model can be degraded to special cases in previous work. (1) Single RIS case, i.e., $ {\bf H}_1 = {\bf H}_3 = {\bf H}_{\rm s} = {\bf 0} $ or $ {\bf H}_2 = {\bf H}_4 = {\bf H}_{\rm s} $. We can obtain the result of single reflection scenario \cite{zhang2020transmitter} or MIMO Rayleigh double scattering channels \cite{hoydis2011iterative}. (2) Two RISs with only single reflections, i.e., $ {\bf H}_{\rm s} = {\bf 0} $. We can obtain the result of $ 2 $ RISs case with single reflection \cite{moustakas2022reconfigurable}.\\
    \end{remark}
    \begin{remark}
    \label{remark2}
    The achievable ergodic rate expression depends only on statistical CSI, i.e., spatial correlations $ {\bf T}_i $ and $ {\bf R}_i $, large scale fading coefficients $ \Gamma_i $, which makes statistical CSI-enabled optimal design feasible.
    \end{remark}

We combine the replacement in (\ref{replacement}) and express $ {\bf Q} $ and $ {\bf \Theta}_i, i \in \{ 1, 2 \} $ explicitly, the large-system asymptotic achievable ergodic rate is obtained as
\begin{align}
\label{new_obj_fun}
& {\bar R}({\bf Q}, {\bf \Theta}_1, {\bf \Theta}_2) = \log \det \left( {\frac{{{\sigma ^2}{\bf{I}} + {e_1}{{\bf{R}}_1} + {e_2}{{\bf{R}}_2}}}{{{\sigma ^2}}}} \right) \\ \notag
& + \log \det \left( {\bf{I}} + {{\tilde e}_1}{\bf{\Theta }}_1^{\rm{H}}{{\bf{T}}_1}{{\bf{\Theta }}_1} ({e_{\rm{s}}}{{\bf{R}}_{\rm{s}}} + e_3 {\bf R}_3) \right)\\ \notag
& + \log \det \left( {{\bf{I}} + \left( {{{\tilde e}_{\rm{s}}}{\bf{\Theta }}_2^{\rm{H}}{{\bf{T}}_{\rm{s}}}{{\bf{\Theta }}_2} + {{\tilde e}_2}{\bf{\Theta }}_2^{\rm{H}}{{\bf{T}}_2}{{\bf{\Theta }}_2}} \right){e_4}{{\bf{R}}_4}} \right) \\ \notag
& + \log \det \left( {{\bf{I}} + \left( {{{\tilde e}_{\rm{3}}}{{\bf{T}}_3} + {{\tilde e}_{\rm{4}}}{{\bf{T}}_4}} \right){\bf{Q}}} \right) \\ \notag
& - {L_1}{e_1}{{\tilde e}_1} - {L_2}{e_2}{{\tilde e}_2} - {L_2}{e_{\rm{s}}}{{\tilde e}_{\rm{s}}} - M{e_3}{{\tilde e}_{\rm{3}}} - M{e_4}{{\tilde e}_{\rm{4}}}.
\end{align}
    \begin{remark}
    \label{remark3}
    One logdet term in (\ref{ergodic_rate_after_replacement}) is decomposed into four logdet terms in (\ref{new_obj_fun}), which are corresponding to the four nodes respectively, i.e., the user, the RIS 1, the RIS 2, and the BS.\\
    \end{remark}
    \begin{remark}
    \label{remark4}
    The transmit beamforming and the double-RIS cooperative beamforming are decoupled since the source covariance matrix $ {\bf Q} $ and the phase-shifting matrices $ {\bf \Theta}_i, i \in \{ 1, 2 \} $ are involved in different $ \log \det $ terms. In another words, the design of the matrices $ {\bf Q} $ and $ {\bf \Theta}_i, i \in \{ 1, 2 \} $ are independent. Specifically, the optimization of $ {\bf Q} $ is only dependent on the $5$th $ \log \det $ term in (\ref{new_obj_fun}), where $ {\bf Q} $ should be aligned with $ {\bf H}_3 $ and $ {\bf H}_4 $ simultaneously.\\
    \end{remark}
    \begin{remark}
    \label{remark5}
    Surprisingly, the designs of $ {\bf \Theta}_1 $ and $ {\bf \Theta}_2 $ are also decoupled since there is no cascaded $ {\bf \Theta}_1 $ and $ {\bf \Theta}_2 $ structure.
    \end{remark}
To obtain the ergodic rate, i.e., the first order moment of rate, the bridge connecting RIS1 and RIS2, i.e., $ {\bf H}_{\rm s} $ is somewhat ``broken" by taking the expectation with respect to $ {\bf W}_{\rm s} $.

With the obtained asymptotic achievable ergodic rate only dependent on long-term channel statistics, the statistical CSI-enabled joint optimal design is conducted in the next section.

\section{Ergodic Rate Optimization}
The asymptotic expression of the ergodic rate of the double-RIS assisted MIMO system in (\ref{new_obj_fun}) makes statistical CSI-enabled optimal design feasible, where the needed statistical CSI can be estimated through the existing channel estimation method \cite{9133156, 9400843, 914500, 5484583}. By substituting (\ref{new_obj_fun}) for the objective function in (P1), we can recast the optimization problem as problem (P2),
    \begin{align}
    \text{(P2)} \quad & \max_{ {\bf Q}, \boldsymbol{\Theta}_1, \boldsymbol{\Theta}_2 } \ \ \bar R\left( {{\bf{Q}}, {\bf{\Theta }}_1, {\bf{\Theta }}_2} \right) \\ \notag
    \text{s.t.} \ \ \ & {\bf \Theta}_i = {\rm diag} \{ \vartheta_{i, 1}, \vartheta_{i, 2}, \cdots, \vartheta_{i, L_i} \}, i \in \{1, 2\}, \\ \notag
    & \lvert \vartheta_{i, l_i} \rvert = 1,  1 \le l_i \le L_i, \\ \notag
    & \tr\left( {\bf Q} \right) \le P, {\bf Q} \succeq 0.
    \end{align}
As we mentioned in {\bf {Remark 4}} and {\bf {Remark 5}}, thanks to our large system analysis, the optimal design of the source covariance matrix $ {\bf Q} $, the phase-shifting matrices $ {\bf \Theta}_1 $ and $ {\bf \Theta}_2 $ are weakly coupled only through the auxiliary variables. Besides, the constraints on the three variables are independent. A natural thought is to optimize the three variables in problem (P2) alternatively.
\subsection{Optimization of ${\bf Q}$ with Fixed ${\bf \Theta}_1$ and ${\bf \Theta}_2$}
By resorting to the AO algorithm, we first obtain the asymptotically optimal source covariance matrix.
By taking the gradient of $ \bar R\left( {{\bf{Q}}} \right) $ with respect to $ {\bf Q} $, we have $ \frac{{\partial \bar R\left( {\bf{Q}} \right)}}{{\partial {{\bf{Q}}}}} = \frac{{\partial \bar R_5\left( {\bf{Q}} \right)}}{{\partial {\bf Q}}} + \sum\limits_i {\frac{{\partial \bar R\left( {\bf{Q}} \right)}}{{\partial {e_i}}}\frac{{\partial {e_i}}}{{\partial {{\bf{Q}}}}}} + \sum\limits_i {\frac{{\partial \bar R\left( {\bf{Q}} \right)}}{{\partial {\tilde e}_i}}\frac{{\partial {\tilde e}_i}}{{\partial {{\bf{Q}}}}}} $ since the source covariance matrix $ {\bf Q} $ is involved in the $ 5 $th term of (\ref{new_obj_fun}), i.e., ${\bar R_5}\left( {\bf{Q}} \right) = \log \det \left( {{{\bf{I}}_M} + {\bf{FQ}}} \right)$, and variables $ e_i, {\tilde e}_i, ~\forall i $.
Obviously, the derivative expression is rather complicated because of the involved circularly defined auxiliary variables $ e_i, {\tilde e}_i ~\forall i $.
However, following the argument in appendix, $ e_i $s and $ {\tilde e}_i $s are calculated through setting the partial derivatives of free energy as zeros. This implies that $ \frac{{\partial \bar R\left( {\bf{Q}} \right)}}{{\partial {e_i}}} = 0, \frac{{\partial \bar R\left( {\bf{Q}} \right)}}{{\partial {\tilde {e}_i}}} = 0 $.
Therefore, the derivative of the objective function only involves the $5$th term of (\ref{new_obj_fun}) where $ {\bf F} = {{{\tilde e}_{\rm{3}}}{{\bf{T}}_3} + {{\tilde e}_{\rm{4}}}{{\bf{T}}_4}} $.
Thus, we simplify the optimization problem as
    \begin{align}
    \label{Q_obj}
    \text{(P2.1)} \quad \max_{ {\bf Q} } \ \ & \log \det \left( {{\bf{I}}_M} + {\bf F} {\bf Q} \right) \\ \notag
    \text{s.t.} \ \ \ & \tr\left( {\bf Q} \right) \le P, {\bf Q} \succeq 0. \notag
    \end{align}
It is found that problem (P2.1) is strictly concave with respect to $ {\bf Q} $, where the Karush-Kuhn-Tucker (KKT) conditions of problem (P2.1) are
\begin{align}
\left\{ \begin{array}{l}
{\bf{F}} ({{\bf{I}}_M} + {\bf{FQ}})^{ - 1} + \iota {\bf{I}}_M - {\bf \Upsilon} = 0,\\
\mu ({\rm{Tr}}({\bf{Q}}) - P) = 0, \iota  \ge 0,\\
{\rm{Tr}}({\bf \Upsilon} {\bf{Q}}) = 0, {\bf \Upsilon} \succeq 0, {\bf{Q}} \succeq 0,
\end{array} \right.
\end{align}
where $ \iota $ and $ {\bf \Upsilon} $ are Lagrange multipliers associated with the inequality constraints. The KKT conditions imply that problem (P2.1) can be solved through a standard water-filling procedure, therefore the following proposition is given to obtain the optimal $ {\bf Q} $.
    \begin{proposition}
    \label{proposition2}
    The expression of optimal source covariance matrix is
    \begin{align}
    \label{Q_opt_WF}
    {\bf Q}^{\rm {opt}} =  {\bf U}_{\bf F} {\bf \Lambda}_{\bf Q} {\bf U}_{\bf F}^{\hem},
    \end{align}
    where the eigenvalue decomposition of matrix $ {\bf F} $ is given by $ {\bf F} = {\bf U}_{\bf F} {\bf \Lambda}_{\bf F} {\bf U}_{\bf F}^{\hem} $. And $ {\bf \Lambda}_{\bf Q} = ( \frac{1}{\iota} {\bf{I}}_M - {\bf \Lambda}_{\bf F}^{-1} )^{+} $, where $ \iota $ is chosen to satisfy the power constraint $ \tr\left( {\bf Q} \right) \le P $.
    \end{proposition}
The optimization of $ {\bf Q} $ is proposed in {\bf{Algorithm 1}}. 
In each step of {\textbf {Algorithm 1}}, the optimization variable $ {\bf Q} $ is calculated through the closed-form solution in (\ref{Q_opt_WF}). And the achievable rate monotonically increases during each iteration. Besides, the value of the achievable rate is upper-bounded by a finite value. Therefore {\textbf {Algorithm 1}} is guaranteed to converge.
\begin{algorithm}[!t]
    \label{Algorithm_1}
    \small
    \centering
    \caption{Given fixed $ {\bf \Theta}_i, i \in \{ 1, 2 \} $ and optimize $ {\bf Q} $}
    \begin{tabular*}{17cm}{l}
    1: \textbf{Initialization:} $ {\bf Q}^{(0)} = {\bf I}_M $, $ e_j^{(0)} = {\tilde e}_j^{(0)} = 1 $, $ ~\forall j $, \\
       \ \ \ \ convergence condition $ \epsilon = 10^{-5} $; \\
    2: Compute $ {\bar R} ( {\bf{Q}}^{(0)}, {\bf{\Theta}}_1, {\bf{\Theta}}_2 ) $ according to (\ref{new_obj_fun}); \\
    3: \textbf{while} $ n = 0 $ or \\
       \ \ \ \ $ | {\bar R} ( {{\bf{Q}}^{(n+1)}, {\bf{\Theta}}_1, {\bf{\Theta}}_2 } ) - {\bar R} ( {{\bf{Q}}^{(n)}, {\bf{\Theta}}_1, {\bf{\Theta}}_2 } ) | > \epsilon $, \textbf{do} \\
    4: \ \ Update $ n = n+1 $; \\
    5: \ \ Compute $ e_j^{(n)} $ {\color{blue}{and}} $ {\tilde e}_j^{(n)} $ according to (\ref{auxiliary_variable_e}); \\
    6: \ \ Compute $ {\bf F} $; \\
    7: \ \ Compute $ {\bf Q}^{(n)} $ according to (\ref{Q_opt_WF}); \\
    8: \ \ Compute $ {\bar R} ( {\bf{Q}}^{(n)}, {\bf{\Theta}}_1, {\bf{\Theta}}_2 ) $ according to (\ref{new_obj_fun}); \\
    9: \ \textbf{end while} \\
    10: \textbf{Output:} Obtain $ {\bf Q}^{\rm {opt}} = {\bf Q}^{(n)} $. \\
   \end{tabular*}
\end{algorithm}

\subsection{Optimization of ${\bf \Theta}_1$ and ${\bf \Theta}_2$ with Fixed ${\bf Q}$}
In this subsection, we optimize the phase-shifting matrices ${\bf \Theta}_1$ and ${\bf \Theta}_2$ with fixed ${\bf Q}$. Therefore, we reformulated the sub-problem as
    \begin{align}
    \text{(P2.2)} \quad & \max_{ \boldsymbol{\Theta}_1, \boldsymbol{\Theta}_2 } \ \ \bar R\left( {{\bf{\Theta }}_1, {\bf{\Theta }}_2} \right) \\ \notag
    \text{s.t.} \ \ \ & {\bf \Theta}_i = {\rm diag} \{ \vartheta_{i, 1}, \vartheta_{i, 2}, \cdots, \vartheta_{i, L_i} \}, i \in \{1, 2\}, \\ \notag
    & \lvert \vartheta_{i, l_i} \rvert = 1,  1 \le l_i \le L_i.
    \end{align}
Different from single-RIS assisted system, the optimization of double-RIS assisted system need to design $2$ phase-shifting matrices, thus there are $3$ variables, i.e., ${\bf \Theta}_1$, ${\bf \Theta}_2$, and $ {\bf Q} $ participating in the AO process. This brings high signal processing complexity for double-RIS system due to the following reasons. First, thanks to the low cost of RIS, large-scale surfaces with hundreds of passive reflecting elements are affordable to implement which makes the dimension of the phase-shifting matrices relatively high. Although our full statistical CSI-enabled optimal design has greatly reduced the signaling overhead compared to instantaneous CSI-enabled design, there are still two large-scale surfaces to be optimized. Second, in our double-RIS system framework, the joint optimal design algorithm is realized by the BS which is responsible for performing most calculation tasks. Specifically, the optimal phase shifts will be transmitted to the RISs through the control link. Then, RISs intelligently adjust the reflecting coefficients through its smart controller. These signaling interactions are complex in our distributed double-RIS systems.
To reduce the signal processing complexity and practical implementation, we consider a common-phase scheme, i.e., the two RISs tune their reflecting elements in a synchronized manner and hold the same phase-shifting matrices $ {\bf \Theta}_1 = {\bf \Theta}_2 = {\bf \Theta} $ and $ L_1 = L_2 = L $ in each time slot
The common-phase scheme achieves a low hardware cost and low energy consumption with a certain level of sacrifice of optimization performance.
With respect to the optimization of $ {\bf \Theta} $, although the expression of new objective function is concise and clear thanks to our large system analysis, the quadratic structure of $ {\bf \Theta} $ involved in the two logdet terms and unit-modulus constraint make the optimization problem still challenging to solve. Besides, the common-phase assumption makes the variables couple tighter. We utilize PSO algorithm to obtain the optimal $ {\bf \Theta} $ owing to its high universality which is easy to implement and has few parameters to adjust \cite{9483943PSO}. Since the implementation of PSO algorithm is standard and similar to {\bf {Algorithm 1}} in \cite{9483943PSO}, we omit the details.

\subsection{Proposed AO Algorithm}
With the optimal source covariance $ {\bf Q}^{\rm {opt}} $ and the optimal phase-shifting matrix $ {\bf{\Theta}}^{\rm {opt}} $, an AO algorithm is proposed to solve problem (P2).
The complete AO algorithm to find $ {\bf Q}^{\rm {opt}} $ and $ {\bf{\Theta}}^{\rm {opt}} $ is presented in {\bf {Algorithm 2}}. Specifically, since common phase shift is considered in our double-RIS design, i.e., $ {\bf \Theta}_1 = {\bf \Theta}_2 = {\bf \Theta} $, the three-tuple alternating process is reduced to a two-variable alternating process.
First, the source covariance is initialized as $ {\bf Q}^{(0)} = {\bf I}_M $, and the phase-shifting matrix $ {\bf{\Theta}}^{(0)} $ is initialized as a random matrix.
Then at each time, one variable is optimized iteratively with the other fixed. Note that the objective function (\ref{Q_obj}) in problem (P2.1) is strictly concave with respect to $ {\bf Q} $, therefore $ {\bf Q}^{\rm {opt}} $ is an optimal solution.
As for $ {\bf \Theta} $, the PSO algorithm will at least find a local optima every time it executes, and the ergodic rate increases monotonically in each iteration. Besides, the objective function is upper-bounded by a finite value which guarantees the convergence of {\bf {Algorithm 2}}.
The complexity of {\bf {Algorithm 2}} is greatly reduced compared to the previous algorithms maximizing capacity. On the one hand, {\bf {Algorithm 2}} designs the system based on statistical CSI only which avoids frequent signaling exchange. On the other hand, since we introduced common-phase scheme, the calculation complexity and signaling overhead are deduced in half.
\begin{algorithm}[!t]
    \label{Algorithm_2}
    \small
    \centering
    \caption{The proposed AO algorithm to slove (P2)}
    \begin{tabular*}{17cm}{l}
    1: \textbf{Initialization:} $ {\bf Q}^{(0)} = {\bf I}_M $. \\
     \ \ \ Randomly generated $ {\bf \Theta}^{(0)} $ with phases $ \{ \theta_l, ~\forall i \} $ following \\
     \ \ \ the independent and uniform distribution in $[0, 2\pi )$, \\
     \ \ \ $ e_j^{(0)} = {\tilde e}_j^{(0)} = 1 $, $ ~\forall j $, convergence condition $ \epsilon = 10^{-5} $; \\
    2: Compute $ {\bar R} ( {\bf{Q}}^{(0)}, {\bf{\Theta}}^{(0)} ) $ according to (\ref{new_obj_fun}); \\
    3: \textbf{while} $ n = 0 $ \\
       \ \ \ \ or $ | {\bar R} ( {{\bf{Q}}^{(n+1)}, {\bf{\Theta}}^{(n+1)} } ) - {\bar R} ( {\bf{Q}}^{(n)}, {\bf{\Theta}}^{(n)} ) | > \epsilon $, \textbf{do} \\
    4: \ \ Update $ n = n+1 $; \\
    5: \ \ Compute $ {\bf Q}^{(n)} $ by {\textbf {Algorithm 1}}; \\
    6: \ \ Compute $ {\bf \Theta}^{(n)} $ according to PSO algorithm; \\
    7: \ \ Compute $ {\bar R} ( {\bf{Q}}^{(n)}, {\bf{\Theta}}^{(n)} ) $ according to (\ref{new_obj_fun}); \\
    8: \textbf{end while}\\
    9: \textbf{Output:} Obtain $ {\bf Q}^{\text{opt}} = {\bf Q}^{(n)} $ and $ {\bf \Theta}^{\text{opt}} = {\bf \Theta}^{(n)} $. \\
   \end{tabular*}
\end{algorithm}

\section{Numerical Results}
In this section, numerical simulations are provided to evaluate the accuracy of our large system analysis results, the efficiency of the proposed optimal design algorithm, and the superiority of the double-RIS assisted MIMO framework. The positions of all the communications nodes are characterized by 3-dimensional spatial locations.
The 3-D Cartesian coordinations of the BS, the user, RIS1, and RIS2 are respectively $ (1, 0, 5) $m, $ (1, 50, 1.5) $m, $ (0, 50, 3) $m, $ (0, 0, 3) $m. Unless otherwise specified, $ 8 $ antennas are equipped at the BS, each RIS is equipped with $ 100 $ reflecting elements, $ 4 $ antennas are equipped at the user. Moreover, the noise power is $-94$dBm, the path loss is a function of distance $ \Gamma_j (d_{TR}) [ dB ] = G_t + G_r -35.1 - 36.7 \log_{10} (d_{TR} / 1 {\rm m}) $, where $ G_t = G_r = 5dBi $ denote the antenna gains at the transmitter and the receiver for each link, respectively \cite{8888223}.
For spatial correlation at the BS and the user, the integral model is adopted \cite{zhang2021large}, where the transmit and receive correlation matrices are generated by $ {\left[ {{\bf{T}}_j\;{\rm{or}}\;{\bf{R}}_j} \right]_{m,n}} = \int\limits_{ - 180}^{180} {\frac{{{\rm{d}}\phi }}{{\sqrt {2\pi {\delta_{{\bf T} / {\bf R}} ^2}} }}}  e^ { {2\pi jd_s ( {m - n} )\sin ( {\frac{{\pi \phi }}{{180}}} ) - \frac{{{{( {\phi  - \eta_{{\bf T} / {\bf R}} } )}^2}}}{{2{\delta ^2}}}} } $, where $ \eta_{{\bf T} / {\bf R}} $ is the mean angle and $ \delta_{{\bf T} / {\bf R}} $ is the angular spread.
In the integral spatial correlation model, $ \phi $ is the antenna array plane's physical angle, which is randomly generated in $ [-\pi, \pi) $.
We set $ { \eta_{\bf T} } = { \eta_{\bf R} } = 0^{\circ} $ as the mean angle, and $ { \delta_{\bf T} } = { \delta_{\bf R} } = 5^{\circ} $ as the angular spread of the radio wave direction cluster. Since RISs are rectangular surfaces, the spatial correlation at the RISs are modeled as $ {\left[ {{\bf{T}}_j\;{\rm{or}}\;{\bf{R}}_j} \right]_{m,n}} = {\rm {sinc}} (2 \| {\bf u}_{m} - {\bf u}_{n} \| / \lambda) $ \cite{9300189}, where $ \| {\bf u}_{m} - {\bf u}_{n} \| $ is the distance between the $m$-th and the $n$-th reflecting elements, and $ \lambda $ is the wavelength.

To validate the accuracy of the obtained expression of the asymptotic ergodic achievable rate, $1000$ Monte-Carlo realizations of instantaneous channel matrices are generated for comparison. Unless otherwise indicated, in the following figures, the solid lines denote the analytical results meanwhile the markers denote the simulation results by Monte-Carlo. For notational brevity, the legend of the analytical results are omitted.
\subsection{Verification of the Large System Analysis}
\begin{figure}[!ht]
    \centering
    \includegraphics[width=0.5\textwidth]{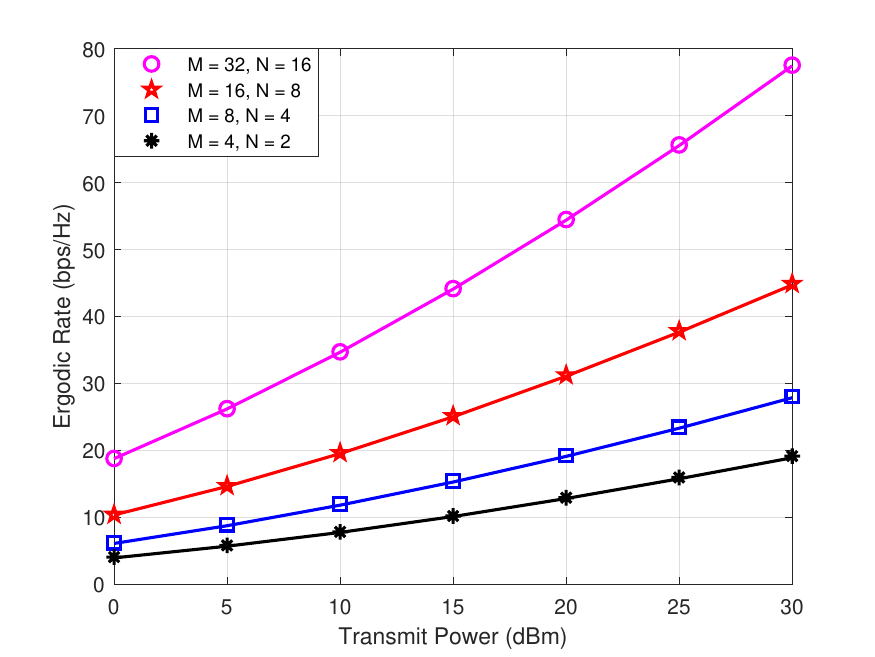}
    \caption{Ergodic rate versus number of transceiver antennas.}
    \label{Anay_antenna}
    \end{figure}

\begin{figure}[!ht]
    \centering
    \includegraphics[width=0.5\textwidth]{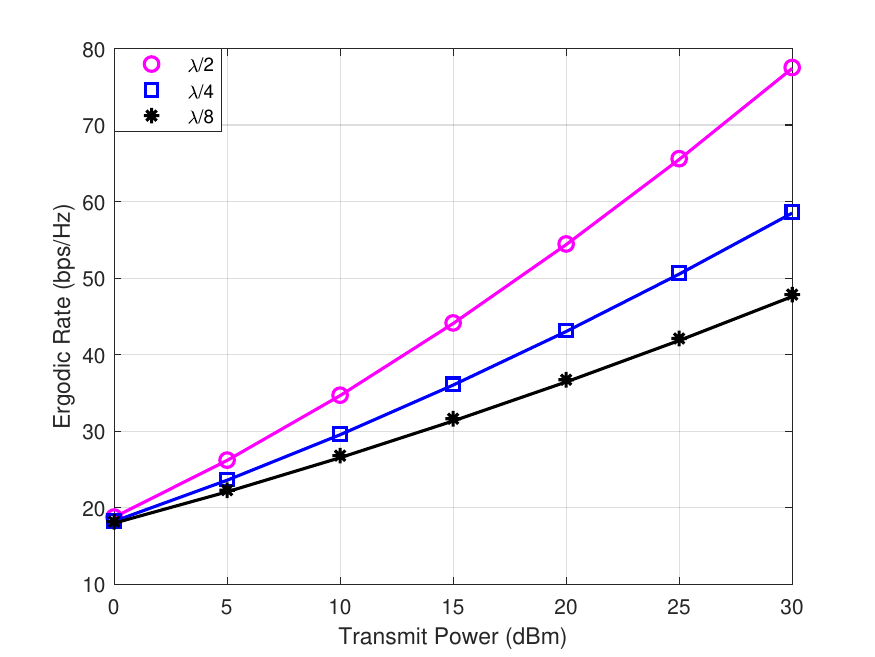}
    \caption{Ergodic rate versus distance of RIS elements.}
    \label{Anay_lambda}
    \end{figure}

Firstly, we investigate the accuracy of the derived asymptotic ergodic rate.
Since we aim to prove the accuracy of our asymptotic result in this subsection, the exact values of the transmit covariance matrix and the phase-shifting matrix do not affect the simulations. Without loss of generality, we set $ {\bf Q} = {\bf I}_M $ and $ {\bf \Theta}_1 = {\bf \Theta}_2 = {\bf I}_L $.
As shown in Fig. \ref{Anay_antenna}, the markers which represent the Monte-Carlo simulation results are nearly directly gone through by the corresponding solid lines representing the analytical results. This phenomenon validates the accuracy of our large system analysis. Particularly, the accuracy is not degraded by the number reduction of antennas at the BS and the user. In another word, the asymptotic approach provides indistinguishable results compared with Monte-Carlo simulations even with a few antennas equipped at the transceiver. This is mainly because of the fast convergence speed advantage of replica method.
    In Fig. \ref{Anay_lambda}, the impact of the reflecting elements spacing is evaluated under massive array scenario, where $ M = 32, N = 16 $. Obviously, RISs with adjacent elements divided by $ \lambda / 2 $ provide larger array aperture which enhances the ergordic rate performance. However, if the elements of RISs are deployed more densely, the spatial correlation will be increased and  DoF of RISs is accordingly decreased, which brings ergodic rate degradation. This discovery matches the results in \cite{9300189}.

\begin{figure}[!ht]
    \centering
    \includegraphics[width=0.5\textwidth]{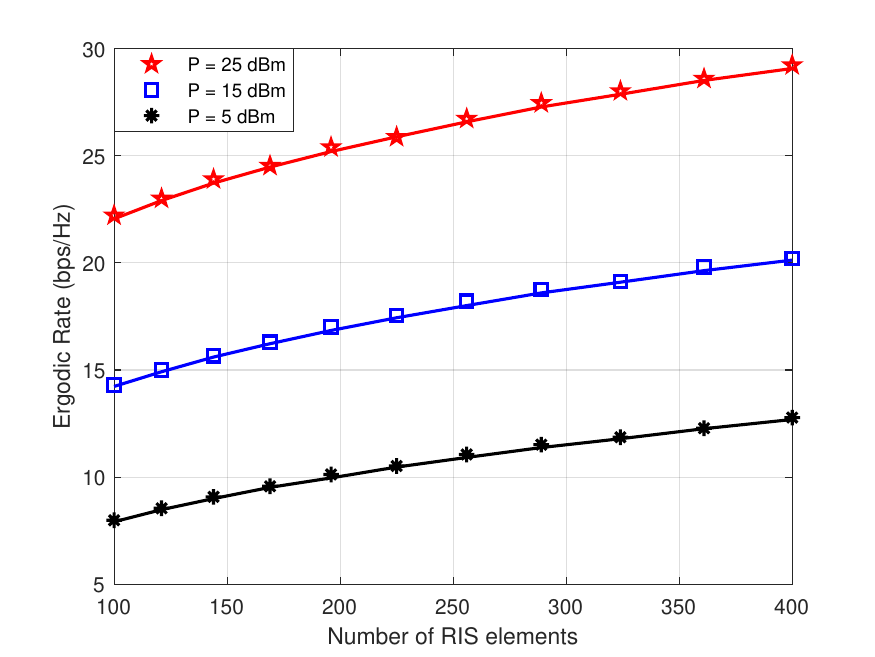}
    \caption{Ergodic rate versus number of RIS elements.}
    \label{Anay_ele}
    \end{figure}

Thanks to the low cost of RIS, large scale surface with hundreds of reflecting elements is affordable. The benefit brought by large surface implementation is depicted in Fig. \ref{Anay_ele}. The ergodic rate increases if we enlarge the scale of reflecting surface due to the increased spatial DoF. It is also found that the marginal effect brought by increasing the elements number becomes small as the surfaces are further enlarged. Therefore, it is not worth the cost of implementing larger RISs to obtain the additional performance gain.
\subsection{Performance of the Proposed AO Algorithm 2}
    \begin{figure}[!ht]
    \centering
    \includegraphics[width=0.5\textwidth]{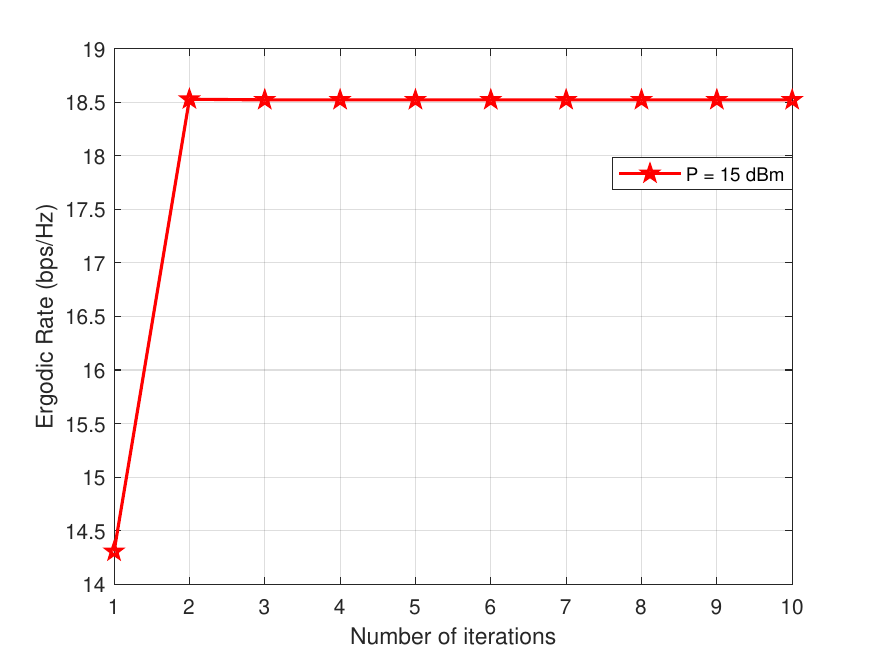}
    \caption{Convergence performance of the proposed algorithm.}
    \label{Conv_analy}
    \end{figure}
In this subsection, the effectiveness of the proposed AO algorithm and common-phase scheme are evaluated.
The convergence performance of the proposed AO algorithm maximizing the achievable ergodic rate of the double-RIS assisted MIMO system by exploiting full statistical CSI is first shown in Fig. \ref{Conv_analy}.
It is observed that {\bf {Algorithm 2}} converges
after only one step which implies its fast convergence rate and computational efficiency of the proposed AO algorithm. The fast convergence speed mainly thanks to the closed-form solution of the precoder, and the carefully chosen parameters for PSO algorithm.
    \begin{figure}[!htpb]
    \centering
    \includegraphics[width=0.5\textwidth]{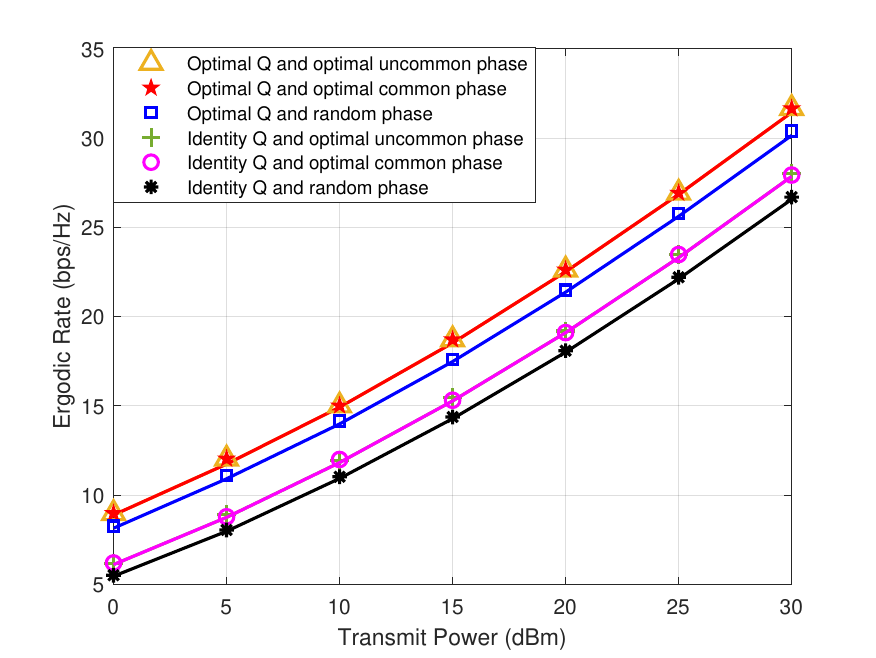}
    \caption{Performance comparison of the proposed optimization algorithm and benchmarks.}
    \label{opt_alg}
    \end{figure}

In Fig. \ref{opt_alg}, the performance of the proposed AO algorithm is investigated compared to the benchmark approaches. It is shown that in high signal-to-noise (SNR) regime, our proposed AO algorithm is able to achieve up to 5 bps/Hz gain.
It is also worth highlighting that optimizing $ {\bf Q} $ only outperforms optimizing phases only, because compared to passive analog beamforming at RIS, digital beamforming at the BS has more advantages.
However, the optimal design of double RISs also brings significant ergodic rate gain, which implies the necessity of jointly optimizing the source covariance matrix at the BS and the phase-shifting matrix of double RISs.
For comparison, Fig. \ref{opt_alg} also demonstrates the performance of uncommon-phase optimization scheme. In uncommon-phase optimization, the phase-shifting matrices $ {\bf \Theta}_1 $ and $ {\bf \Theta}_2 $ of RIS1 and RIS2 are designed separately. In another words, the two variables $ {\bf \Theta}_1 $ and $ {\bf \Theta}_2 $ participate in the alternating process when optimizing the passive RISs phases. It is shown that under both initial ${\bf Q}$ and optimal ${\bf Q}$ the performances of optimal common phase and optimal uncommon phase are nearly the same in Fig. \ref{opt_alg}.
In practice, the performance of common-phase design is a lower bound of the uncommon-phase design. In our simulation, the tightness of the lower bound is demonstrated. This is because under the topology of the simulation setup, the optimal phase-shifting matrices of RIS1 and RIS2 are nearly equal to each other.
Therefore, only half of the complexity and signaling overhead is cost while maintaining a superior performance under our common-phase scheme.
\subsection{Effectiveness of the Proposed Double-RIS Assisted MIMO Framework}
    \begin{figure}[!htpb]
    \centering
    \includegraphics[width=0.5\textwidth]{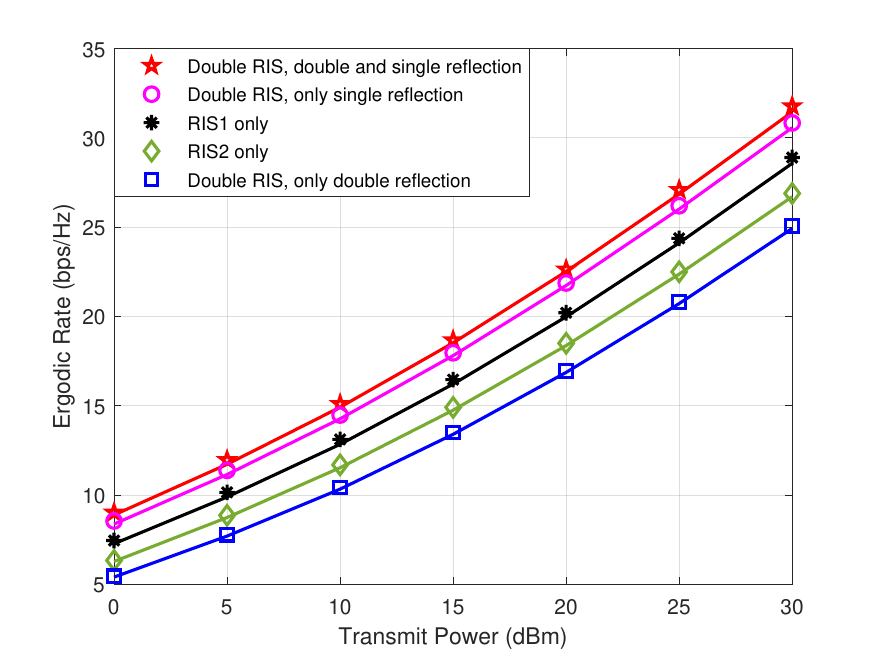}
    \caption{Performance comparison of different RIS deployments.}
    \label{deployments}
    \end{figure}
Finally, we investigate the performance of double-RIS framework compared with RIS-assisted MIMO with only single reflection. To evaluate the contribution of cooperative passive beamforming through secondary reflection, double-RIS with only double reflection is also chosen as a benchmark. Thus the four benchmarks are: (1) Double RISs with only single reflection, i.e., $ {\bf H}_{\rm s} = {\bf 0} $. (2) RIS1 only, i.e., $ {\bf H}_2 = {\bf 0}, {\bf H}_4 = {\bf 0} $, and $ {\bf H}_{\rm s} = {\bf 0} $. (3) RIS2 only, i.e., $ {\bf H}_1 = {\bf 0}, {\bf H}_3 = {\bf 0} $, and $ {\bf H}_{\rm s} = {\bf 0} $. (4) Double RISs with only double reflection, i.e., $ {\bf H}_2 = {\bf 0}, {\bf H}_3 = {\bf 0} $. It is observed in Fig. \ref{deployments} double RISs with both single and double reflection achieves higher ergodic rate over the four benchmark frameworks. Intuitively, the achievable ergodic rate under this framework is contributed by both the single and double reflection. Compare to RIS1-only scenario and RIS2-only scenario, the double-RISs with both single and double reflection scenario performs superiorly at the same cost, since we utilized common-phase optimization scheme.
Moreover, double RISs with only single reflection scenario performs closely to double RISs with both single and double reflection scenario, however, the inter-RIS cooperative beamforming also contributes substantially to the achievable ergodic rate, which can be proven by the performance of double RISs with only double reflection scenario. Although the last benchmark approach, i.e., double RISs with only double reflection performs the worse compared to others due to the high multiplicative path loss, the BS-RIS2-RIS1-user link offers an extra link DoF for reliability.
Once the single reflection link, i.e., $ {\bf H}_2 $ and $ {\bf H}_3 $ are blocked by the undesirable factors in the harsh communication environment, the double reflection link is able to realize path diversity to support a certain amount of data rate. This validates the necessity of the consideration of the double-RIS framework to enhance the reliability performance.

\section{Conclusion}
In this paper, we studied a double-RIS assisted MIMO wireless communication system under spatially correlated Rayleigh fading channel where both single and double reflection links were considered.
To analyze the achievable ergodic rate with complicated matrices sum and product structure in the effective channel, RMT method was utilized assuming that the antennas size tends to infinity.
The derived closed-form large-system approximation of the ergodic rate also perfectly fit with the Monte Carlo simulation results even when deploying a small number of antennas.
With this asymptotic expression, the joint optimal design maximizing the achievable ergodic rate based on full statistical CSI becomes feasible.
By alternatively designing the source covariance matrix and the passive beamforming matrices of the double RISs, we proposed an AO algorithm with common-phase scheme where the double RISs holds the equal phase-shifting matrices in a synchronized manner. The proposed AO algorithm with common-phase scheme outperforms the benchmark schemes with low pilot training overhead and low complexity. And the investigated double-RIS assisted MIMO framework not only achieves better ergodic rate performance compared with RIS-assisted MIMO system with only single reflection, but also has high path diversity due to the inter-RIS cooperation beamforming.

\section*{ACKNOWLEDGEMENT}
\label{ACKNOWLEDGEMENT}

The work of Jiajia Guo was supported in part by the Guangdong Basic and Applied Basic Research Foundation under Grant 2023A1515110732.
The work of Jun Zhang was supported in part by the National Natural Science Foundation of China (NSFC) under Grant 62071247. 
The work of Shaodan Ma was supported in part by the Science and Technology Development Fund, Macau SAR under Grants 0087/2022/AFJ and 001/2024/SKL, in part by the National Natural Science Foundation of China under Grant 62261160650, and in part by the Research Committee of University of Macau under Grants MYRG-GRG2023-00116-FST-UMDF and MYRG2020-00095-FST.
The work of Shi Jin was supported in part by the NSFC under Grant 62261160576 and 62301148, and in part by the Fundamental Research Funds for the Central Universities under Grant 2242023K5003.

\appendix
In the appendix, we prove {{\bf Proposition 1}}.
\subsection{Mutual Information}
We conduct the proof of \emph{Proposition 1} starting from the mutual information definition \cite{Cover2006} which is the origin of the representation of the ergodic achievable rate (\ref{R}).
Assume that the probability distribution functions (PDF) of the channel $ {\boldsymbol {\mathcal H}} = \lbrace {\bf H}_j, ~\forall j \rbrace $ and the input signal $ p( {\bf x} ) $ are given$^1$,
{\footnotetext[1]{At the receiver side, the signal can be detected by Bayesian estimation by using the Bayes formula $p({\bf x}|{\bf y}, {\boldsymbol {\mathcal H}}) = \frac{p({\bf y}|{\bf x}, {\boldsymbol {\mathcal H}})p({\bf x})}{p({\bf y}|{\boldsymbol {\mathcal H}})}$.}}
for the double-RIS based MIMO channel, the conditional mutual information is given by
        \begin{align}
    \label{mutual_information_def}
    & {\emph I} \ ({\bf x} ; {\bf y} | {\boldsymbol {\mathcal H}}) = \mathbb{E}_{ \lbrace {\bf x}, {\bf y} \rbrace }
    {\lbrace \log p( {\bf y}|{\bf x}, {\boldsymbol {\mathcal H}} ) \Big | \boldsymbol {\mathcal H} \rbrace} \\ \notag
    & - \mathbb{E}_{ \lbrace {\bf x}, {\bf y} \rbrace }
    {\lbrace \log { p( {\bf y} | {\boldsymbol {\mathcal H}} ) } \Big | \boldsymbol {\mathcal H} \rbrace}.
    \end{align}
We have the conditional PDF of the output signals using the Gaussian property of the noise $ \bf n $. 
    \begin{align}
    \label{pdf_y_cond}
    p({\bf y}|{\bf x}, {\boldsymbol {\mathcal H}}) = \frac{1}{(\pi \sigma^2)^N}e^{-\frac{1}{\sigma^2} \| {\bf y} - {\bf H}_{\rm {eff}} {\bf x} \|^2}.
    \end{align}
As for the first iterm of (\ref{mutual_information_def}), we obtain $ \mathbb{E}_{ \lbrace {\bf x}, {\bf y} \rbrace } \lbrace \log p( {\bf y}|{\bf x}, {\boldsymbol {\mathcal H}} ) | \boldsymbol {\mathcal H} \rbrace = -N $ by substituting (\ref{pdf_y_cond}).
As for the second iterm of (\ref{mutual_information_def}), the conditional PDF of the received signal is $ p( {\bf y}|{\boldsymbol {\mathcal H}} ) = \mathbb{E}_{\bf x} [ p( {\bf y} | {\bf x}, {\boldsymbol {\mathcal H}} ) ] $. Therefore, the overall information is given by
    \begin{align}
    \label{mutual_information}
    {\emph I} \ ( {\bf x} ; {\bf y} ) & = \mathbb{E}_{ {\boldsymbol {\mathcal H}} }
    {\lbrace {\emph I} \ ( {\bf x} ; {\bf y} | {\boldsymbol {\mathcal H}} ) \rbrace} \\ \notag
    & = -\mathbb{E}_{ \lbrace {\bf y}, {\boldsymbol {\mathcal H}} \rbrace }
    {\lbrace \log \mathbb{E}_{\bf x} \lbrace e^{ -\frac {1} {\sigma^2} \| {\bf y} - {\bf H}_{\rm {eff}} {\bf x} \|^2 } \rbrace \rbrace} - N.
    \end{align}
We first define the significant partition function in this complex system as
    \begin{align}
    \label{TDef}
    T ( {\bf y}, {\boldsymbol {\mathcal H}} ) \triangleq \mathbb{E}_{\bf x} \lbrace e^{ -\frac {1} {\sigma^2} \| {\bf y} - {\bf H}_{\rm {eff}} {\bf x} \|^2 } \rbrace.
    \end{align}
Then the free energy is expressed as
    \begin{align}
    F \triangleq -\mathbb{E}_{ \lbrace {\bf y}, {\boldsymbol {\mathcal H}} \rbrace }
    {\lbrace \log T ( {\bf y}, {\boldsymbol {\mathcal H}} ) \rbrace}.
    \label{FreeEnDef}
    \end{align}
Now the bridge of achievable rate and free energy is built, which are essentially the information entropy and thermodynamic entropy respectively.

It is observed from (\ref{FreeEnDef}) that the most challenging task is the high-dimensional integral of a logarithm function, where the integral variables are random matrices $ {\boldsymbol {\mathcal H}} $ and $ {\bf y} $.
The replica method is then utilized to conduct the analysis by the property $ {\lim_{r \to 0}} \ \frac{\partial} {\partial r} \log \mathbb{E} \lbrace X^r \rbrace = {\lim_{r \to 0}} \frac{\mathbb{E} \lbrace X^r \log X \rbrace} {\mathbb{E} \lbrace X^r \rbrace} = \mathbb{E} \lbrace \log X \rbrace $ for any positive random variable $ X $. Then the free energy is re-expressed as
    \begin{align}
    \label{FreeEn_replica}
    F = -{\lim_{r \to 0}} \ \frac{\partial} {\partial r} \log \mathbb{E}_{ \lbrace {\bf y}, {\boldsymbol {\mathcal H}} \rbrace } \lbrace T^r (  {\bf y}, {\boldsymbol {\mathcal H}} ) \rbrace.
    \end{align}
Through replica method, we can calculate the asymptotic free energy $ \mathcal F = {\lim_{M \to \infty}} F $ in a large system regime, where $ M $ tends to infinity.
Replica method firstly requires the replica of the vector of transmit signal. Here we define for $ \alpha = 0, 1, 2, \cdots, r $, the $ \alpha $-th replica of $ {\bf x} $ as $ {\bf x}^{(\alpha)} $ with $ p({\bf x}) $ as their input PDF.
Then combining the all $ r+1 $ replicas of $ {\bf x} $, the replica set of the transmit signal is $ {\bf{X}} \triangleq [ {\bf x}^{(0)}, {\bf x}^{(1)}, \cdots, {\bf x}^{(r)} ] \in \mathbb{C}^{M \times (r+1)} $.
Substituting the partition function (\ref{TDef}) into the free energy (\ref{FreeEn_replica}), the logarithmic part is expressed as
    \begin{align}
    \label{logTerm}
    & \mathbb{E}_{ \lbrace {\bf y}, {\boldsymbol {\mathcal H}} \rbrace } \lbrace T^r ( {\bf y}, {\boldsymbol {\mathcal H}} ) \rbrace
    = \int \mathbb{E}_{\boldsymbol {\mathcal H}} \lbrace p( {\bf y} | {\boldsymbol {\mathcal H}} ) T^r ( {\bf y}, {\boldsymbol {\mathcal H}} ) \rbrace \mathrm{d} {\bf y} \\ \notag
    & = \mathbb{E}_{ \lbrace {\bf X}, {\boldsymbol {\mathcal H}} \rbrace } \lbrace \int \mathrm{d} {\bf y} \frac{1}{(\pi \sigma^2)^N} \prod_{\alpha = 0}^r e^{ -\frac {1} {\sigma^2} \| {\bf y} - {\bf H}_{\rm {eff}} {\bf x}^{(\alpha)} \|^2 } \rbrace.
    \end{align}
Since $ {\bf H}_{\rm {eff}} = {{\bf{H}}_{\rm{r}}}{{{\bf{\mathord{\buildrel{\lower3pt\hbox{$\scriptscriptstyle\frown$}}
\over H} }}}_{\rm{s}}}{{\bf{H}}_{\rm{t}}} $, we deal with the random components matrix-by-matrix.
\subsection{Expectation Over Random Channels}
In this subsection, the $ 3 $ random matrices involved in the effective channel are integrated with respect to their random components respectively. The basic idea is to deal with one random matrix at a time while fixing others as deterministic matrices. In our double-RIS assisted MIMO system, the double reflection and single reflection are both considered, therefore the integral order should be properly chosen and the replica sets corresponding to the random channels should be delicately designed.
\subsubsection{Expectation Over $ {\bf H}_{\rm r} $}
Fixing ${{{\bf{\mathord{\buildrel{\lower3pt\hbox{$\scriptscriptstyle\frown$}} \over H} }}}_{\rm{s}}}$ and $ {{\bf{H}}_{\rm{t}}} $ as constants, take the expectation over the random component in $ {\bf H}_{\rm r} $.
The replica set corresponding to $ {\bf H}_{\rm r} $ is defined as $ {\bf V}_{\rm{r}} \triangleq [ {\bf v}_{\rm{r}}^{( 0 )}, {\bf v}_{\rm{r}}^{( 1 )}, \cdots, {\bf v}_{\rm{r}}^{( r )} ] = ({{\bf{H}}_1}\left( {{{\bf{H}}_3} + {{\bf{H}}_{\rm{s}}}{{\bf{H}}_4}} \right) + {{\bf{H}}_2}{{\bf{H}}_4}) {\bf X}$, where $ {\bf v}_{\rm r}^{( \alpha )} \triangleq ( {{\bf{H}}_1}\left( {{{\bf{H}}_3} + {{\bf{H}}_{\rm{s}}}{{\bf{H}}_4}} \right) + {{\bf{H}}_2}{{\bf{H}}_4} ) {\bf x}^{( \alpha )}, ~\forall \alpha $.
As $ M \to \infty $, $ {\bf V}_{\rm r} $ converges to a zero-mean Gaussian random matrix according to the central limit theorem, and its covariance is $ {\bf C}_{\rm r} = {\bf C}_{ {\rm r} 1} \otimes {\bf R}_1 + {\bf C}_{ {\rm r} 2} \otimes {\bf R}_2 $, where $ \{ {\bf C}_{ {\rm r} 1}, {\bf C}_{ {\rm r} 2} \} \in \mathbb{C}^{ (r+1) \times (r+1) } $ are matrices with entries given by $ [ {\bf C}_{ {\rm r} 1} ]_{ \alpha, \beta } = \frac{1}{L_1} ( ( {{{\bf{H}}_3} + {{\bf{H}}_{\rm{s}}}{{\bf{H}}_4}} ) {\bf x}^{( \alpha )} )^\hem {\bf T}_1 ( ( {{{\bf{H}}_3} + {{\bf{H}}_{\rm{s}}}{{\bf{H}}_4}} ) {\bf x}^{( \beta )} ) $, and $ [ {\bf C}_{ {\rm r} 2} ]_{ \alpha, \beta } = \frac{1}{L_2} ( {\bf H}_4 {\bf x}^{( \alpha )} )^\hem {\bf T}_2 ( {\bf H}_4 {\bf x}^{( \beta )} ) $, for $ \alpha, \beta = 0, \cdots, r $. Then (\ref{logTerm}) can be re-expressed as a function including probability measures of $ {\bf C}_{\rm r} $ according to large deviation theory,
    \begin{align}
    \label{logTerm_measure12}
    & \mathbb{E}_{ \lbrace {\bf y}, {\boldsymbol {\mathcal H}} \rbrace } \lbrace T^r (  {\bf y}, {\boldsymbol {\mathcal H}} ) \rbrace \\ \notag
    & = \mathbb{E}_{\bf X} \lbrace \int e^{ {\mathcal G}_{\rm r}^{(r)} ({\bf C}_{\rm r}) } \mathrm{d} {\mu}_{\rm r}^{(r)} ({\bf C}_{\rm r}) \rbrace + {\mathcal O}(1),
    \end{align}
where $ {\mathcal G}_{\rm r}^{(r)} ({\bf C}_{\rm r}) $ and $ {\mu}_{\rm r}^{(r)} ({\bf C}_{\rm r}) $ are the probability measures as represented below.
    \begin{align}
    \label{measureG12}
    & {\mathcal G}_{\rm r}^{(r)} ({\bf C}_{\rm r}) = \log \mathbb{E}_{{\bf V}_{\rm r}} \lbrace \int \mathrm{d} {\bf y} \frac{1}{(\pi \sigma^2)^N} \prod_{\alpha = 0}^r e^{-\frac{1}{\sigma^2} \| {\bf y} - {\bf v}_{\rm r}^{(\alpha)} \|^2} \rbrace \\ \notag
    & = \log \mathbb{E}_{{\bf V}_{\rm r}} \lbrace \int \mathrm{d} {\bf y} \frac{1}{(\pi \sigma^2)^N} e^{-\frac{1}{\sigma^2} \sum_{\alpha = 0}^{r} \| {\bf y} - {\bf v}_{\rm r}^{(\alpha)} \|^2} \rbrace \\ \notag
    & \overset {(a)} {=} \log \mathbb{E}_{{\bf V}_{\rm r}} \lbrace e^{ -\tr( { {\bf V}_{\rm r} {\bf \Sigma} {\bf V}_{\rm r}^\hem } )} \rbrace - N \log (1+r) \\ \notag
    & \overset {(b)} {=} - \log \det {\bf D}_{\rm r} - N \log (1+r),
    \end{align}
where $ {{\bf{D}}_{\rm{r}}} = {\bf{I}} + \left( {{\bf{\Sigma }} \otimes {\bf{I}}} \right){{\bf{C}}_{\rm{r}}} $ and $ {\bf \Sigma} \triangleq \frac{1}{\sigma^2 (r+1)} ( (r+1){\bf I}_{r+1} - {\bf 1}_{r+1} {\bf 1}_{r+1}^\mathrm{T} ) $. In (\ref{measureG12}), Gaussian integral and Hubbard-Stratonovich transformation is used for $ (a) $.
The property of Kronecker product is used in (\ref{measureG12}) for $ (b) $.
And we can obtain the measure $ \mu^{(r)}_{\rm r} ({\bf C}_{\rm r}) $ as
    \begin{align}
    \label{measure_mu_12}
    & \mu^{(r)}_{\rm r} ({\bf C}_{\rm r}) = \\ \notag
    & \prod_{\alpha, \beta = 0}^r \delta ( ( ( {{{\bf{H}}_3} + {{\bf{H}}_{\rm{s}}}{{\bf{H}}_4}} ) {\bf x}^{( \alpha )} )^\hem {\bf T}_1 ( ( {{{\bf{H}}_3} + {{\bf{H}}_{\rm{s}}}{{\bf{H}}_4}} ) {\bf x}^{( \beta )} ) \\ \notag
    & - L_1 [ {\bf C}_{{\rm r} 1} ]_{ \alpha, \beta } ) \delta ( ( {\bf H}_4 {\bf x}^{( \alpha )} )^\hem {\bf T}_2 ( {\bf H}_4 {\bf x}^{( \beta )} ) - L_2 [ {\bf C}_{{\rm r} 2} ]_{ \alpha, \beta } ) \\ \notag
    & = e^{ -{\mathcal R}_{\rm r}^{(r)} ({\bf C}_{\rm r}) + {\mathcal O}(1) },
    \end{align}
where $ {\mathcal O}(\cdot) $ denotes a constant as $ M \to \infty $, and $ \delta(\cdot) $ is the Dirac function. $ {\mathcal R}_{\rm r}^{(r)} ( {\bf C}_{\rm r} ) = {\mathcal R}_{{\rm r} 1}^{(r)} ( {\bf C}_{{\rm r} 1} ) + {\mathcal R}_{{\rm r} 2}^{(r)} ( {\bf C}_{{\rm r} 2} ) $ is the rate measure of $ {\mu}^{(r)}_{\rm r} ({\bf C}_{\rm r}) $ which is given by
    \begin{subequations}
    \label{measure_R_12}
    \begin{align}
    \label{measure_R_1}
    & {\mathcal R}_{{\rm r} 1}^{(r)} ( {\bf C}_{{\rm r} 1} ) = \max_{\tilde {\bf C}_{{\rm r} 1}} \lbrace L_1 \tr ( {\tilde {\bf C}_{{\rm r} 1}} {\bf C}_{{\rm r} 1} ) \\ \notag
    & - \log \mathbb{E}_{ \lbrace {\bf X}, {\boldsymbol {\mathcal H}} \rbrace } \lbrace e^{ \tr ( {\tilde {\bf C}_{{\rm r} 1}} {\bf X}^\hem ( {{{\bf{H}}_3} + {{\bf{H}}_{\rm{s}}}{{\bf{H}}_4}} )^\hem {\bf T}_1 ( {{{\bf{H}}_3} + {{\bf{H}}_{\rm{s}}}{{\bf{H}}_4}} ) {\bf X} ) } \rbrace \rbrace, \\
    \label{measure_R_2}
    & {\mathcal R}_{{\rm r} 2}^{(r)} ( {\bf C}_{{\rm r} 2} ) = \max_{\tilde {\bf C}_{{\rm r} 2}} \lbrace L_2 \tr ( {\tilde {\bf C}_{{\rm r} 2}} {\bf C}_{{\rm r} 2} ) \\ \notag
    & - \log \mathbb{E}_{ \lbrace {\bf X}, {\boldsymbol {\mathcal H}} \rbrace } \lbrace e^{ \tr ( {\tilde {\bf C}_{{\rm r} 2}} {\bf X}^\hem {\bf H}_4^\hem {\bf T}_2 {\bf H}_4 {\bf X} ) } \rbrace \rbrace,
    \end{align}
    \end{subequations}
with $ \{ {\tilde {\bf C}}_{{\rm r} 1}, {\tilde {\bf C}}_{{\rm r} 1} \} \in \mathbb{C}^{(r+1) \times (r+1)} $ being symmetric matrices.
Substituting (\ref{measure_mu_12}) into (\ref{logTerm_measure12}), we have $ \mathbb{E}_{ \lbrace {\bf y}, {\boldsymbol {\mathcal H}} \rbrace } \lbrace T^r ( {\bf y}, {\boldsymbol {\mathcal H}} ) \rbrace = \mathbb{E}_{ \lbrace {\bf X}, {\boldsymbol {\mathcal H}} \rbrace } \lbrace \int e^{ {\mathcal G}_{\rm r}^{(r)} ({\bf C}_{\rm r}) - {\mathcal R}_{\rm r}^{(r)} ({\bf C}_{\rm r}) + {\mathcal O}(1) } \mathrm{d} {\bf C}_{\rm r} \rbrace $. Therefore, the integration with respect to $ {\bf C}_{\rm r} $ can be computed through the saddle point method as $ M \to \infty $,
    \begin{align}
    \label{saddle12}
    & -{\lim_{M \to \infty}} \log \mathbb{E}_{ \lbrace {\bf y}, {\boldsymbol {\mathcal H}} \rbrace } \lbrace T^r ( {\bf y}, {\boldsymbol {\mathcal H}} ) \rbrace \\ \notag
    & = \min_{ {\bf C}_{\rm r} } \lbrace - {\mathcal G}_{\rm r}^{(r)} ({\bf C}_{\rm r}) + {\mathcal R}_{\rm r}^{(r)} ({\bf C}_{\rm r}) \rbrace.
    \end{align}
\subsubsection{Expectation Over $ {\bf H}_{\rm s} $}
Fixing $ {\bf H}_{\rm r} $ and $ {{\bf{H}}_{\rm{t}}} $ as constants, take the expectation over the random component in $ {\bf H}_{\rm s} $.
The corresponding replicated random vectors are defined as $ {\bf v}_{\rm s}^{( \alpha )} \triangleq {\bf H}_{\rm s} {\bf H}_4 {\bf x}^{( \alpha )}, ~\forall \alpha $, and the replica set $ {\bf V}_{\rm s} \triangleq {\bf H}_{\rm s} {\bf H}_4 {\bf X} = [ {\bf v}_{\rm s}^{( 0 )}, {\bf v}_{\rm s}^{( 1 )}, \cdots, {\bf v}_{\rm s}^{( r )} ] $. As before, from the generalized central limit theorem, the random matrix $ {\bf V}_{\rm s} $ converges to a Gaussian random matrix, with zero mean and covariance $ {\bf C}_{\rm s} \otimes {\bf R}_{\rm s} $, where $ {\bf C}_{\rm s} \in \mathbb{C}^{ (r+1) \times (r+1) } $ is a matrix with entries given by $ [ {\bf C}_{\rm s} ]_{ \alpha, \beta } = \frac{1}{L_2} ( {\bf H}_4 {\bf x}^{( \alpha )} )^\hem {\bf T}_{\rm s} ( {\bf H}_4 {\bf x}^{( \beta )} ) $, for $ \alpha, \beta = 0, \cdots, r $. Based on (\ref{saddle12}), the exponent terms which contain random matrices $ {\bf H}_{\rm s} $, $ {\bf H}_3 $, $ {\bf H}_4 $ and $ {\bf X} $ in (\ref{measureG12}) and (\ref{measure_R_12}) can be reformulated as
    \begin{align}
    \label{exp_term_remain_s}
    & \mathbb{E}_{ \lbrace {\bf X}, {\boldsymbol {\mathcal H}} \rbrace } \lbrace e^{ \tr ( {{{\bf{\tilde C}}}_{{\rm{r}}1}}{{\bf{X}}^{\rm{H}}}{({{\bf{H}}_3} + {{\bf{H}}_{\rm{s}}}{{\bf{H}}_4})^{\rm{H}}}{{\bf{T}}_1}({{\bf{H}}_3} + {{\bf{H}}_{\rm{s}}}{{\bf{H}}_4}){\bf{X}} ) } \rbrace \\ \notag
    & = \mathbb{E}_{ \lbrace {\bf X}, {\boldsymbol {\mathcal H}} \rbrace } \lbrace \int e^{ {\mathcal G}_{\rm s}^{(r)} ({\bf C}_{\rm s}) } \mathrm{d} \mu_{\rm s}^{(r)} ({\bf C}_{\rm s}) \rbrace + {\mathcal O}(1),
    \end{align}
where the measure $ { {\mathcal G}_{\rm s}^{(r)} ({\bf C}_{\rm s}) } $ is expressed as
    \begin{align}
    \label{measure_G_s}
    { {\mathcal G}_{\rm s}^{(r)} ({\bf C}_{\rm s}) } = \mathbb{E}_{ \{ {\boldsymbol {\mathcal H}}, {\bf X} \} } \{ \tr ( {\bf J}_{\rm s} ) \} - \log \det {\bf D}_{\rm s},
    \end{align}
where ${\bf J}_{\rm s} = {\bf{H}}_3^{\rm{H}}{{\bf{N}}_{\rm{s}}}{{\bf{H}}_3}{\bf{X}}{{\bf{M}}_{\rm{s}}}{{\bf{X}}^{\rm{H}}}$, $ {{\bf{M}}_{\rm{s}}} \otimes {{\bf{N}}_{\rm{s}}} = {{\bf{E}}_{{\rm{ss}}}}\left( {{{\bf{A}}_{\rm{s}}} \otimes {{\bf{B}}_{\rm{s}}}} \right){{\bf{E}}_{{\rm{ss}}}} + {{\bf{E}}_{{\rm{ss}}}} $, $ {{\bf{A}}_{\rm{s}}} \otimes {{\bf{B}}_{\rm{s}}} = ({{\bf{C}}_{\rm{s}}} \otimes {{\bf{R}}_{\rm{s}}}){\bf{D}}_{\rm{s}}^{ - 1} $, $ {\bf D}_{\rm s} = {\bf{I}} - {{\bf{E}}_{{\rm{ss}}}}({{\bf{C}}_{\rm{s}}} \otimes {{\bf{R}}_{\rm{s}}}) $, $ {{\bf{E}}_{{\rm{ss}}}} = {{\bf{\tilde C}}_1} \otimes {{\bf{T}}_1} $. Note that the sign in $ {\bf D}_{\rm s} $ is different from that in $ {\bf D}_{\rm r} $ but will be the same as that in $ {\bf D}_{\rm t} $. And the measure $ \mu^{(r)}_{\rm s} ({\bf C}_{\rm s}) $ is expressed as
    \begin{align}
    \label{measure_mu_s}
    \mu^{(r)}_{\rm s} ({\bf C}_{\rm s}) & = \prod_{\alpha, \beta = 0}^r \delta ( ( {\bf H}_4 {\bf x}^{( \alpha )} )^\hem {\bf T}_{\rm s} ( {\bf H}_4 {\bf x}^{( \beta )} ) - L_2 [ {\bf C}_{\rm s} ]_{ \alpha, \beta } ) \\ \notag
    & = e^{ -{\mathcal R}_{\rm s}^{(r)} ({\bf C}_{\rm s}) + {\mathcal O}(1) },
    \end{align}
where $ {\mathcal R}_{\rm s}^{(r)} $ is the rate measure of $ {\mu}^{(r)}_{\rm s} ({\bf C}_{\rm s}) $ and is given by
    \begin{align}
    \label{measure_R_s}
    & {\mathcal R}_{\rm s}^{(r)} ({\bf C}_{\rm s}) = \max_{\tilde {\bf C}_{\rm s}} \lbrace L_2 \tr ( {\tilde {\bf C}_{\rm s}} {\bf C}_{\rm s} ) \\ \notag
    & - \log \mathbb{E}_{ \lbrace {\bf X}, {\boldsymbol {\mathcal H}} \rbrace } \lbrace e^{ \tr ( {\tilde {\bf C}_{\rm s}} {\bf X}^\hem {\bf H}_4^\hem {\bf T}_{\rm s} {\bf H}_4 {\bf X} ) } \rbrace \rbrace,
    \end{align}
with $ {\tilde {\bf C}}_{\rm s} \in \mathbb{C}^{(r+1) \times (r+1)} $ being a symmetric matrix. Substitute (\ref{measure_mu_s}) into (\ref{exp_term_remain_s}), we have

$ \int e^{ {\mathcal G}_{\rm s}^{(r)} ({\bf C}_{\rm s}) - {\mathcal R}_{\rm s}^{(r)} ({\bf C}_{\rm s}) + {\mathcal O}(1) } \mathrm{d} {\bf C}_{\rm s} $. Therefore, the integration with respect to $ {\bf C}_{\rm s} $ can be calculated through the saddle point method as $ M \to \infty $,
    \begin{align}
    \label{saddle_s}
    & -{\lim_{M \to \infty}} \log \mathbb{E}_{ \lbrace {\bf y}, {\boldsymbol {\mathcal H}} \rbrace } \lbrace T^r ( {\bf y}, {\boldsymbol {\mathcal H}} ) \rbrace \\ \notag
    & = \min_{ {\bf C}_{\rm s} } \lbrace - {\mathcal G}_{\rm s}^{(r)} ({\bf C}_{\rm s}) + {\mathcal R}_{\rm s}^{(r)} ({\bf C}_{\rm s}) \rbrace.
    \end{align}

\subsubsection{Expectation Over $ {\bf H}_{\rm t} $}
Fixing $ {\bf H}_{\rm r} $ and $ {{\bf{H}}_{\rm{s}}} $ as constants, take the integral with respect to the random component in $ {\bf H}_{\rm t} $.
The corresponding random vectors are defined as $ {\bf v}_{3}^{( \alpha )} \triangleq {\bf H}_3 {\bf x}^{( \alpha )}, {\bf v}_{4}^{( \alpha )} \triangleq {\bf H}_4 {\bf x}^{( \alpha )}, ~\forall \alpha $, and the replica set $     {\bf V}_{3} \triangleq {\bf H}_3 {\bf X} = [ {\bf v}_{3}^{( 0 )}, {\bf v}_{3}^{( 1 )}, \cdots, {\bf v}_{3}^{( r )} ], {\bf V}_{4} \triangleq {\bf H}_4 {\bf X} = [ {\bf v}_{4}^{( 0 )}, {\bf v}_{4}^{( 1 )}, \cdots, {\bf v}_{4}^{( r )} ] $.
The random matrices $ {\bf V}_{3} $ and $ {\bf V}_{4} $ also converges to a Gaussian random matrices, with zero means and the covariances $ {\bf C}_{3} \otimes {\bf R}_3 $, $ {\bf C}_{4} \otimes {\bf R}_4 $ where $ \{ {\bf C}_{3}, {\bf C}_{4} \} \in \mathbb{C}^{ (r+1) \times (r+1) } $ are matrices with entries given by $ [ {\bf C}_{3} ]_{ \alpha, \beta } = \frac{1}{M} ( {\bf x}^{( \alpha )} )^\hem {\bf T}_3 ( {\bf x}^{( \beta )} ), [ {\bf C}_{4} ]_{ \alpha, \beta } = \frac{1}{M} ( {\bf x}^{( \alpha )} )^\hem {\bf T}_4 ( {\bf x}^{( \beta )} ) $, for $ \alpha, \beta = 0, \cdots, r $.
Based on (\ref{saddle_s}), the exponenential terms which contain random matrices $ {\bf H}_3 $ and $ {\bf H}_4 $ can be represented as
    \begin{align}
    \label{exp_term_remain_t}
    & \mathbb{E}_{ \lbrace {\bf X}, {\boldsymbol {\mathcal H}} \rbrace } \lbrace e^{ {\rm EXP}_{\rm t} } \rbrace \\ \notag
    & = \mathbb{E}_{ \lbrace {\bf X}, {\boldsymbol {\mathcal H}} \rbrace } \lbrace \int e^{ {\mathcal G}_{{\rm t}}^{(r)} ({\bf C}_{{\rm t}}) } \mathrm{d} \mu_{{\rm t}}^{(r)} ({\bf C}_{{\rm t}}) \rbrace + {\mathcal O}(1),
    \end{align}
where $ { {\rm EXP}_{\rm t} } = {\rm{Tr}}({\bf{H}}_3^{\rm{H}}{{\bf{N}}_{\rm{s}}}{{\bf{H}}_3}{\bf{X}}{{\bf{M}}_{\rm{s}}}{{\bf{X}}^{\rm{H}}}) + {\rm{Tr}}({{{\bf{\tilde C}}}_{\rm{s}}}{{\bf{X}}^{\rm{H}}}{\bf{H}}_4^{\rm{H}}{{\bf{T}}_{\rm{s}}}{{\bf{H}}_4}{\bf{X}}) + {\rm{Tr}}({{{\bf{\tilde C}}}_{{\rm{r2}}}}{{\bf{X}}^{\rm{H}}}{\bf{H}}_4^{\rm{H}}{{\bf{T}}_2}{{\bf{H}}_4}{\bf{X}}) $,  the measure $ { {\mathcal G}_{{\rm t}}^{(r)} ({\bf C}_{{\rm t}}) } $ is expressed as
    \begin{align}
    \label{measure_G_t}
    { {\mathcal G}_{{\rm t}}^{(r)} ({\bf C}_{{\rm t}}) } = - \log \det {\bf D}_{3} - \log \det {\bf D}_{4},
    \end{align}
in which $ {{\bf{D}}_3} = {\bf{I}} - {{\bf{E}}_{33}}\left( {{{\bf{C}}_3} \otimes {{\bf{R}}_3}} \right) $, $ {{\bf{D}}_{\rm{4}}} = {\bf{I}} - {{\bf{E}}_{{\rm{44}}}}\left( {{{\bf{C}}_{\rm{4}}} \otimes {{\bf{R}}_4}} \right) $, $ {{\bf{E}}_{33}} = {{\bf{M}}_{\rm{s}}} \otimes {{\bf{N}}_{\rm{s}}} $, $ {{\bf{E}}_{44}} = {{\bf{\tilde C}}_{\rm{s}}} \otimes {{\bf{T}}_{\rm{s}}} + {{\bf{\tilde C}}_{{\rm{r2}}}} \otimes {{\bf{T}}_2} $. And the measure $ \mu^{(r)}_{{\rm t}} ({\bf C}_{{\rm t}}) $ is expressed as
    \begin{align}
    \label{measure_mu_t}
    & \mu^{(r)}_{{\rm t}} ({\bf C}_{{\rm t}}) = \prod_{\alpha, \beta = 0}^r \delta ( ( {\bf x}^{( \alpha )} )^\hem {\bf T}_3 ( {\bf x}^{( \beta )} ) - M [ {\bf C}_3 ]_{ \alpha, \beta } ) \\ \notag
    & \times \delta ( ( {\bf x}^{( \alpha )} )^\hem {\bf T}_4 ( {\bf x}^{( \beta )} ) - M [ {\bf C}_4 ]_{ \alpha, \beta } ) \\ \notag
    & = e^{ -{\mathcal R}_{{\rm t}}^{(r)} ({\bf C}_{{\rm t}}) + {\mathcal O}(1) },
    \end{align}
where $ {\mathcal R}_{{\rm t}}^{(r)} ( {\bf C}_{3}, {\bf C}_{4} ) $ is the rate measure of $ {\mu}^{(r)}_{{\rm t}} ({\bf C}_{{\rm t}}) $ and is given by
    \begin{align}
    \label{measure_R_t}
    & {\mathcal R}_{{\rm t}}^{(r)} ( {\bf C}_{{\rm t}} ) = \\ \notag
    & \max_{\tilde {\bf C}_{3}} \lbrace M \tr ( {\tilde {\bf C}_{3}} {\bf C}_{3} ) - \log \mathbb{E}_{ \lbrace {\bf X} \rbrace } \lbrace e^{ \tr ( {\tilde {\bf C}_{3}} {\bf X}^\hem {\bf T}_3 {\bf X} ) } \rbrace \rbrace \\ \notag
    & + \max_{\tilde {\bf C}_{4}} \lbrace M \tr ( {\tilde {\bf C}_{4}} {\bf C}_{4} ) - \log \mathbb{E}_{ \lbrace {\bf X} \rbrace } \lbrace e^{ \tr ( {\tilde {\bf C}_{4}} {\bf X}^\hem {\bf T}_4 {\bf X} ) } \rbrace \rbrace,
    \end{align}
with $ \{ {\tilde {\bf C}}_{3}, {\tilde {\bf C}}_{4} \} \in \mathbb{C}^{(r+1) \times (r+1)} $ being symmetric matrices.

Substituting (\ref{measure_mu_t}) into (\ref{exp_term_remain_t}), we have $ \int e^{ {\mathcal G}_{{\rm t}}^{(r)} ({\bf C}_{{\rm t}}) - {\mathcal R}_{{\rm t}}^{(r)} ({\bf C}_{{\rm t}}) + {\mathcal O}(1) } \mathrm{d} {\bf C}_{{\rm t}} $. Therefore, the integration with respect to $ {\bf C}_{{\rm t}} $ can be calculated through the saddle point method as $ M \to \infty $,
    \begin{align}
    \label{saddle_t}
    & -{\lim_{M \to \infty}} \log \mathbb{E}_{ \lbrace {\bf y}, {\boldsymbol {\mathcal H}} \rbrace } \lbrace T^r ( {\bf y}, {\boldsymbol {\mathcal H}} ) \rbrace \\ \notag
    & = \min_{ {\bf C}_{{\rm t}} } \lbrace - {\mathcal G}_{{\rm t}}^{(r)} ({\bf C}_{{\rm t}}) + {\mathcal R}_{{\rm t}}^{(r)} ({\bf C}_{{\rm t}}) \rbrace.
    \end{align}
\subsubsection{Expectation Over $ {\bf X} $}
From (\ref{measure_R_t}), the only random matrix left is $ {\bf X} $. We calculate $ - \log \mathbb{E}_{ \lbrace {\bf X} \rbrace } \lbrace e^{ {\rm{Tr(}}{{{\bf{\tilde C}}}_3}{{\bf{X}}^{\rm{H}}}{{\bf{T}}_3}{\bf{X}}{\rm{)}} + {\rm{Tr(}}{{{\bf{\tilde C}}}_{\rm{4}}}{{\bf{X}}^{\rm{H}}}{{\bf{T}}_4}{\bf{X}}{\rm{)}} } \rbrace = \log \det ({\bf{I}} - {{\bf{E}}_{{\bf{X}}}}) $, where ${{\bf{E}}_{\bf{X}}} = {{\bf{\tilde C}}_3} \otimes {{\bf{T}}_3} + {{\bf{\tilde C}}_{\rm{4}}} \otimes {{\bf{T}}_4}$.
\subsubsection{Replica Symmetry, Free Energy Expression, and Auxiliary Scalar Variables}
The final steps are similar to those of the proof in \cite{9530675, zhang2021large}. Due to the space limitation, the details of these steps are omitted. Then {\emph {Proposition 1}} is proven.

\bibliographystyle{gbt7714-numerical}
\bibliography{myref}


\begin{CCJNLbiography}{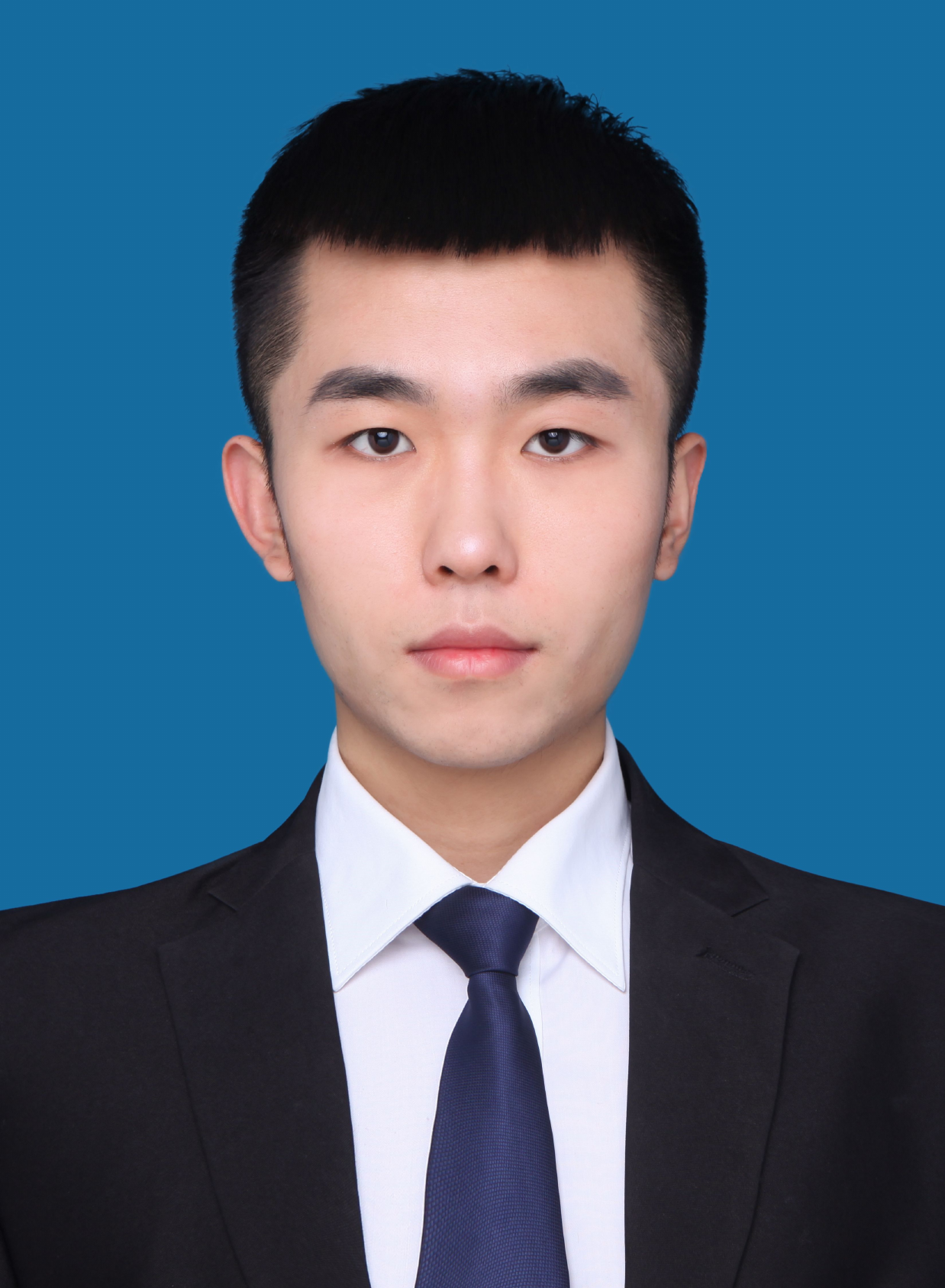}{Kaizhe Xu}
(S’20-M'24) received the B.Eng. and M.Eng. degrees from Jilin University, Changchun, China, in 2016 and 2019 respectively. And he received the Ph.D. degree from the State Key Laboratory of Internet of Things for Smart City, University of Macau, Macau, China in 2023.
He is currently serving as an Assistant Professor in the the Department of Communications and Networking, School of Advanced Technology at Xi'an Jiaotong-Liverpool University.
His current research interest is multi-antenna technology for 6G communications, including reconfigurable intelligent surface, large system analysis for MIMO system, and optimal transceiver design.
\end{CCJNLbiography}

\begin{CCJNLbiography}{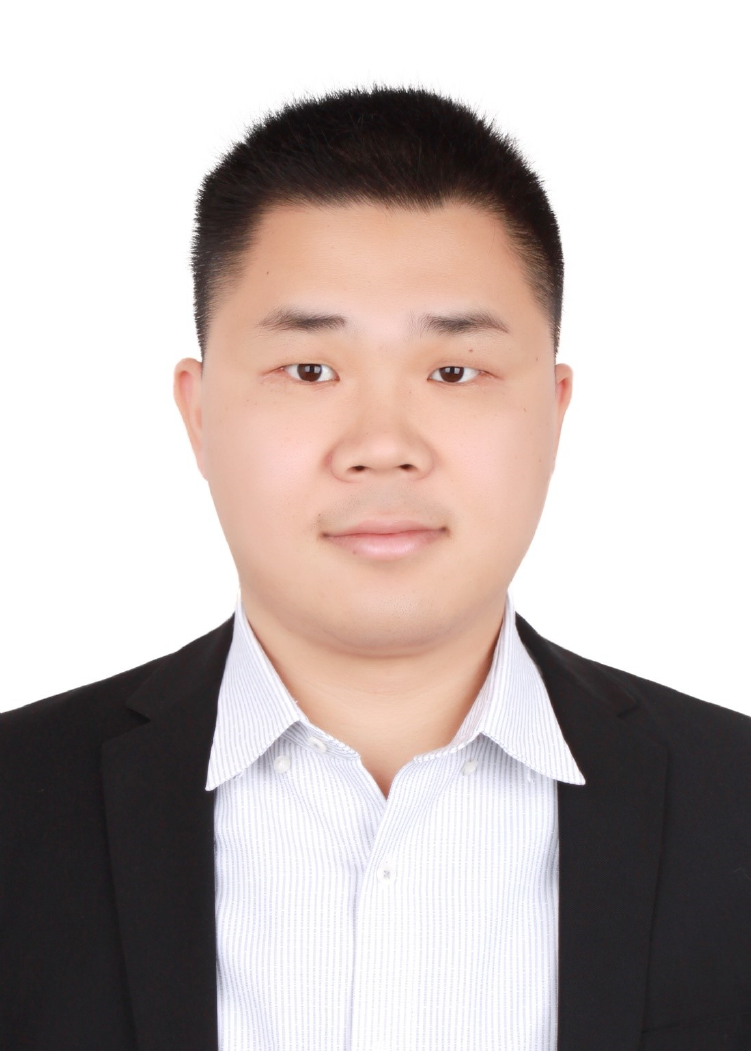}{Jiajia Guo}
(S’21-M'23) received the B.S. degree from the Nanjing University of Science and Technology, Nanjing, China, in 2016, the M.S. degree from the University of Science and Technology of China, Hefei, China, in 2019, and the Ph.D. degree in information and communications engineering from Southeast University, Nanjing, China, in 2023. His current research interests include AI-native air interface, RIS, and massive MIMO. His achievements were selected as one of the Top 10 Science and Technology Progress in the Information and Communication field for 2022 in China. He was the recipient of the First Prize from Natural Science of the Chinese Institute of Electronics and the 2023 Chinese Institute of Electronics Best Doctoral Thesis Award.
\end{CCJNLbiography}

\begin{CCJNLbiography}{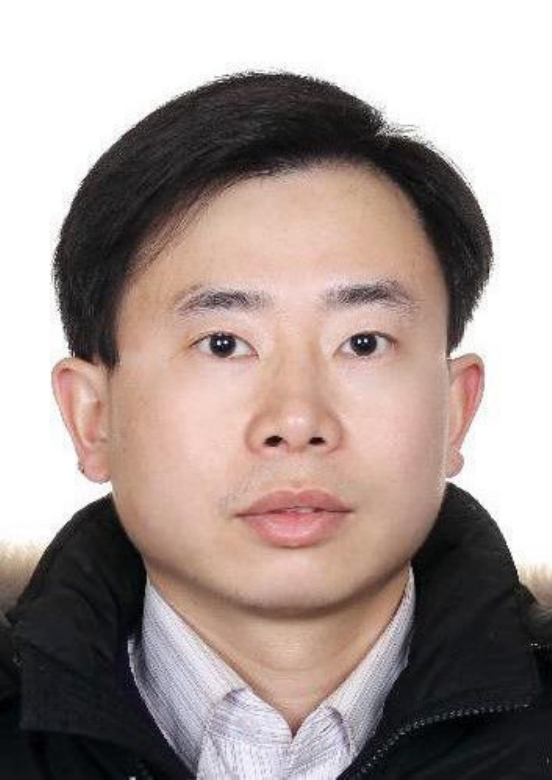}{Jun Zhang}
(S’10–M’14–SM’21) received the M.S. degree in Statistics with Department of Mathematics from Southeast University, Nanjing, China, in 2009, and the Ph.D. degree in Communications Information System with the National Mobile Communications Research Laboratory, Southeast University, Nanjing, China, in 2013. From 2013 to 2015, he was a Postdoctoral Research Fellow with Singapore University of Technology and Design, Singapore. Since 2015, he is with the faculty of the Jiangsu Key Laboratory of Wireless Communications, College of Telecommunications and Information Engineering, Nanjing University of Posts and Telecommunications, where he is currently a Professor. His research interests include massive MIMO communications, RIS-assisted wireless communications, UAV-assisted wireless communications, physical layer security, and large dimensional random matrix theory. Dr. Zhang was a recipient of the Globcom Best Paper Award in 2016, the IEEE APCC Best Paper Award in 2017, the IEEE JC\&S Best Paper Award in 2022, the IEEE/CIC ICCC Best Paper Award in 2023, and the WCSP Best Paper Award in 2023. He has served as an Associate Editor for the IEEE Communications Letters.    
\end{CCJNLbiography}

\begin{CCJNLbiography}{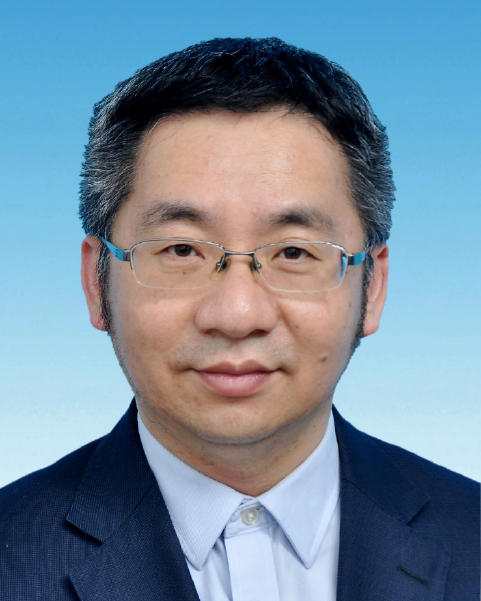}{Shi Jin}
(S’06-M’07-SM'17-F’24) received the B.S. degree in communications engineering from Guilin University of Electronic Technology, Guilin, China, in 1996, the M.S. degree from Nanjing University of Posts and Telecommunications, Nanjing, China, in 2003, and the Ph.D. degree in information and communications engineering from the Southeast University, Nanjing, in 2007. From June 2007 to October 2009, he was a Research Fellow with the Adastral Park Research Campus, University College London, London, U.K. He is currently with the faculty of the National Mobile Communications Research Laboratory, Southeast University. His research interests include wireless communications, random matrix theory, and information theory. He is serving as an Area Editor for the Transactions on Communications and IET Electronics Letters. He was an Associate Editor for the IEEE Transactions on Wireless Communications, and IEEE Communications Letters, and IET Communications. Dr. Jin and his co-authors have been awarded the 2011 IEEE Communications Society Stephen O. Rice Prize Paper Award in the field of communication theory, the IEEE Vehicular Technology Society 2023 Jack Neubauer Memorial Award, a 2022 Best Paper Award and a 2010 Young Author Best Paper Award by the IEEE Signal Processing Society.
\end{CCJNLbiography}

\begin{CCJNLbiography}{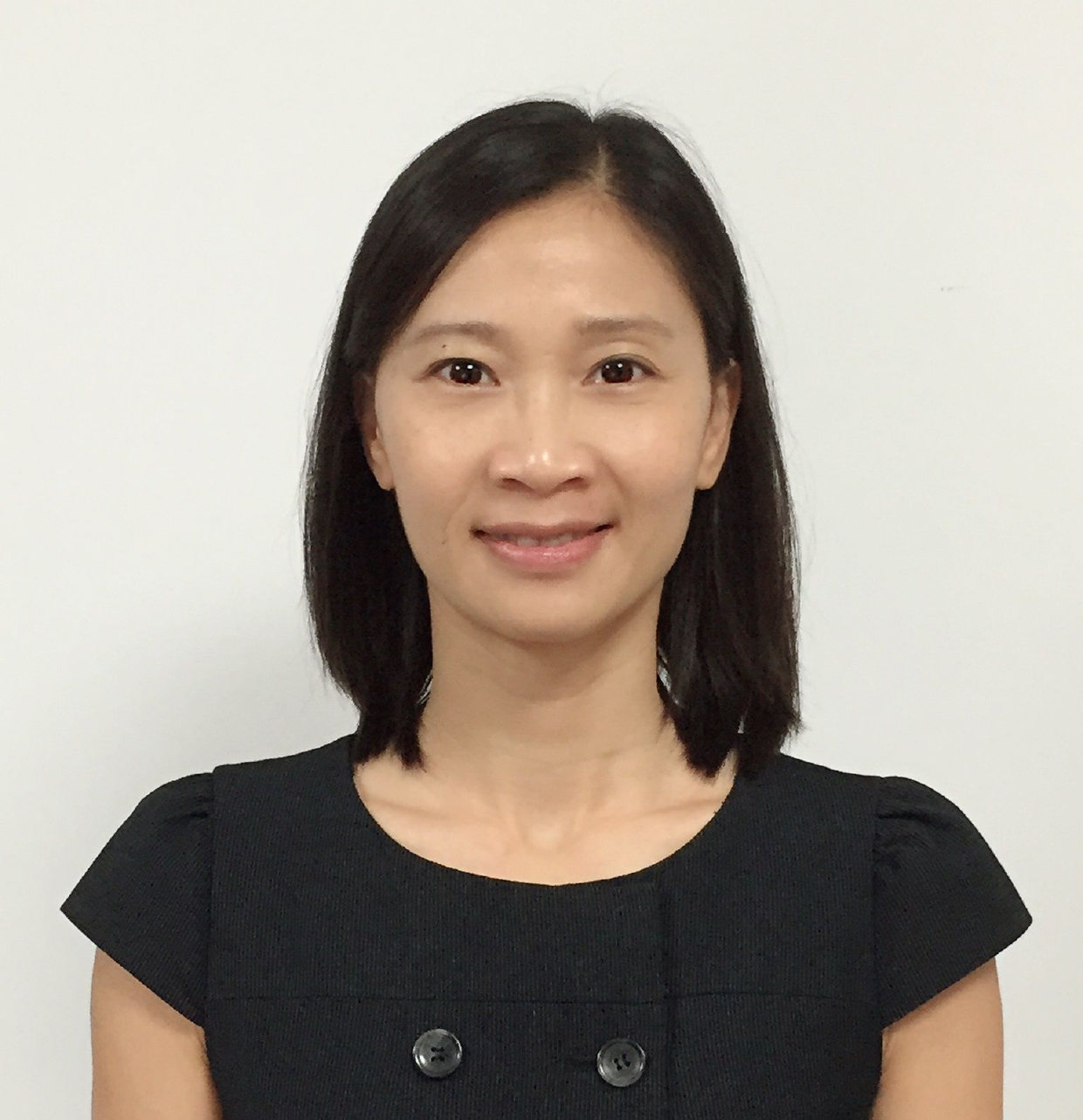}{Shaodan Ma}
received the double Bachelor’s degrees in science and economics and the M.Eng. degree in electronic engineering from Nankai University, Tianjin, China, in 1999 and 2002, respectively, and the Ph.D. degree in electrical and electronic engineering from The University of Hong Kong, Hong Kong, in 2006. From 2006 to 2011, she was a post-doctoral fellow at The University of Hong Kong. Since August 2011, she has been with the University of Macau, where she is currently a Professor. Her research interests include array signal processing, transceiver design, localization, integrated sensing and communication, mmwave communications, massive MIMO, and machine learning for communications. She was a symposium co-chair for various conferences including IEEE ICC 2021, 2019 \& 2016, IEEE GLOBECOM 2016, IEEE/CIC ICCC 2019, etc. She has served as an Editor for IEEE Wireless Communications (2024-present), IEEE Communications Letters (2023), Journal of Communications and Information Networks (2021-present), IEEE Transactions on Wireless Communications (2018-2023), IEEE Transactions on Communications (2018-2023), and IEEE Wireless Communications Letters (2017-2022).
\end{CCJNLbiography}

\end{document}